\documentclass[iop]{emulateapj}
\slugcomment{{\sc Accepted to ApJ:} November 27, 2012} 
\usepackage{graphicx}
\usepackage{epstopdf}
\usepackage{apjfonts}
\usepackage{natbib}
\usepackage{rotating}
\begin{document}
\title{The Clustering of Extremely Red Objects}
\author{David P. Palamara$^{1,2}$, Michael J. I. Brown$^{1,2}$, Buell T. Jannuzi$^{3,4}$, Arjun Dey$^3$, Daniel Stern$^5$, Kevin A. Pimbblet$^{1,2}$, Benjamin J. Weiner$^4$, Matthew L. N. Ashby$^6$, C. S. Kochanek$^{7,8}$, Anthony Gonzalez$^9$, Mark Brodwin$^{10}$, Emeric Le Floc'h$^{11}$ and Marcia Rieke$^{4}$}
\affil{1. School of Physics, Monash University, Clayton, Victoria 3800, Australia}
\affil{2. Monash Center for Astrophysics (MoCA), Monash University, Clayton, Victoria 3800, Australia}
\affil{3. National Optical Astronomy Observatory, 950 N. Cherry Ave., Tucson, AZ 85719, USA}
\affil{4. Steward Observatory, University of Arizona, 933 N. Cherry Ave., Tucson, AZ, 85721}
\affil{5. Jet Propulsion Laboratory, California Institute of Technology, Pasadena, CA 91109, USA}
\affil{6. Harvard-Smithsonian Center for Astrophysics, 60 Garden Street, Cambridge, MA 02138, USA}
\affil{7. Department of Astronomy, The Ohio State University, 140, West 18th Avenue, Columbus, OH 43210, USA}
\affil{8. Center for Cosmology and Astroparticle Physics, The Ohio State University, 191 W. Woodruff Avenue, Columbus, OH 43210, USA}
\affil{9. Department of Astronomy, University of Florida, Gainesville, FL 32611-2055, USA}
\affil{10. Department of Physics and Astronomy, University of Missouri, 5110 Rockhill Road, Kansas City, MO 64110}
\affil{11. Laboratoire AIM, CEA-Saclay -- CNRS -- Universite Paris Diderot, Service d'Astrophysique, Orme des Merisiers, 91191 Gif-sur-Yvette, France}

%abstract - updated 27/11/12
\begin{abstract}
We measure the clustering of Extremely Red Objects (EROs) in $\approx 8~{\rm{deg}^2}$ of the NOAO Deep Wide Field Survey Bo\"otes field in order to establish robust links between ERO ($z\approx1.2$) and local galaxy ($z<0.1$) populations. Three different color selection criteria from the literature are analyzed to assess the consequences of using different criteria for selecting EROs. Specifically, our samples are $(R-K_s)>5.0$ ($28,724$ galaxies), $(I-K_s)>4.0$ ($22,451$ galaxies) and $(I-[3.6])>5.0$ ($64,370$ galaxies). Magnitude-limited samples show the correlation length ($r_0$) to increase for more luminous EROs, implying a correlation with stellar mass. We can separate star-forming and passive ERO populations using the $(K_s-[24])$ and $([3.6]-[24])$ colors to $K_s=18.4$ and $[3.6]=17.5$, respectively. Star-forming and passive EROs in magnitude limited samples have different clustering properties and host dark halo masses, and cannot be simply understood as a single population. Based on the clustering, we find that bright passive EROs are the likely progenitors of $\gtrsim4L^*$ elliptical galaxies. Bright EROs with ongoing star formation were found to occupy denser environments than star-forming galaxies in the local Universe, making these the likely progenitors of $\gtrsim L^*$ local ellipticals. This suggests that the progenitors of massive $\gtrsim4L^*$ local ellipticals had stopped forming stars by $z\gtrsim1.2$, but that the progenitors of less massive ellipticals (down to $L^*$) can still show significant star formation at this epoch.
\end{abstract}

%Introduction - updated 27/11/12
\section{Introduction}
\label{intro}
The pioneering $K$-band near-infrared survey of \citet{elsto88} was the first to identify galaxies with red colors of $(R-K)>5$, and hypothesised these to be either a new class of galaxy at $z>6$, or ellipticals at $z>1$. \citet{graha96} named these galaxies Extremely Red Objects (EROs) for their extreme optical to infrared (IR) colors (e.g., $(R-K)_{Vega}>5.0$, $(I-K)_{Vega}>4.0$, $(R-[3.6])_{Vega}>6.5$) and recorded the first redshift for an ERO of $z=1.44$. We now know that the class of EROs consists of two $z>0.8$ sub-populations: galaxies with old red stellar populations \citep[e.g.,][]{spinr97} or star-forming galaxies with strong dust-obscuration \citep[e.g., from interaction and/or orientation;][]{dey99,yan03}.

Since the highest mass galaxies at this redshift are often found to be EROs \citep[$\gtrsim10^{11.5}~\rm{M}_{\scriptsize{\odot}}$; e.g.,][]{moust04,conse08}, and because EROs also exhibit very strong clustering, comparable to the local clustering of massive elliptical galaxies \cite[e.g.,][]{daddi00,daddi02,brown05}, EROs have been posited as the progenitors of massive local ellipticals. This is important because the formation history of massive galaxies provides a crucial test of galaxy formation models.  A key part of any such test is to link high and low redshift galaxy populations, and with the advent of large-area IR surveys, EROs provide an ideal and easily accessible testing ground for this purpose \citep[][]{gonza08}.

One powerful means of linking galaxy populations across redshift is through clustering and the (dark matter) halo occupation model \citep[e.g.,][]{moust02}. In the case of massive dark matter halos, simulations of dark matter structure evolution such as the Millennium Run \citep[][]{sprin05} demonstrate that massive halos evolve rapidly in mass but little in (comoving) space over cosmic time.  Since the most massive galaxies occupy the most massive halos \citep[e.g.,][]{selja00}, they too grow in mass but their (comoving) spatial distribution evolves slowly. This means that accurate measurements of ERO spatial clustering and host halo mass can be used to link these massive $z\gtrsim0.8$ galaxies with local galaxy populations. 

Previous studies have all found that EROs are very strongly clustered \citep[e.g.,][]{daddi00,firth02,roche02,daddi02,brown05,georg05,kong06,kong09,kim11}. However, the existing estimates of the clustering strengths, as measured by the scale $r_0$ where the two-point correlation strength is unity, show a broad range ($r_0=5.5-17h^{-1}$Mpc) that makes it difficult to link the higher redshift EROs with local galaxy populations. Many of these discrepancies are likely due to differing selection criteria and cosmic variance due to small survey areas and sample sizes. Differences in methodology between studies, such as redshift distribution models and uncertainty estimates, are also possible contributors. We discuss the consequences of these factors below.

The diversity in ERO selection criteria are largely due to pragmatic choices for selecting high-redshift galaxy populations in different data sets. The present range of ERO selection criteria include (but are not limited to): $(R-K)_{\rm{Vega}}>5.0$ \citep[e.g.,][]{brown05}, $(I-K)_{\rm{Vega}}>4.0$ \citep[e.g.,][]{kong09}, $(R-H)_{\rm{Vega}}>4.0$ \citep[e.g.,][]{firth02} and $(R-[3.6])_{\rm{Vega}}>6.5$ \citep[e.g.][]{wilso04}. The consequences of using different ERO selection criteria have been considered in some studies \citep[e.g.,][]{yan03,wilso04,conse08,kim11}. \citet{conse08} found that $\approx60\%$ of EROs were selected by both $(I-K)$ and $(R-K)$ criteria, with the main difference being a shift in the mean redshift ($\bar{z}$). Similarly, \citet{kim11} attribute the difference in clustering between $(r-K)$ and $(i-K)$ selected EROs to be a consequence of their difference in $\bar{z}$. \citet{wilso04} replaced the $K$-band with IRAC $3.6\mu$m, and suggested that ERO samples selected by this color will include more dusty EROs with star formation than $K-$band selected EROs (as we will discuss in \S\ref{separation}). 

Morphological and spectroscopic studies of EROs show them to be a mixture of galaxies with and without star formation \citep[e.g.,][]{smail02,moust04,stern06,wilso07,conse08,kong09}. The relative fractions of dusty galaxies with active star formation (hereafter, star-forming), and galaxies dominated by old stellar populations and little ongoing star formation (hereafter, passive) depends on the ERO selection criteria and the limiting magnitude \citep[e.g.,][]{moust04,conse08}. The sub-populations can be very broad, as star-forming EROs can have star formation rates (SFRs) of between $10-200~\rm{M}_{\scriptsize{\odot}}/yr$ \citep[e.g.,][]{messi10}, but distinguishing further between sub-populations of star-forming EROs is beyond the scope of this study. Separating passive from star-forming EROs is important because we should not expect the two ERO types to possess the same stellar mass, occupy the same mass dark matter halos or be the progenitors of the same local galaxy populations.

Different fractions of these two ERO sub-populations selected by different ERO criteria may well explain the large differences seen in ERO clustering studies. Unfortunately, accurate morphological and spectroscopic classifications are often unavailable for the sources used in ERO clustering studies. In most cases, colors are used to distinguish between star-forming and passive EROs. NIR colors \citep[e.g., ($J-K$);][]{pozze00,kim11} cannot adequately distinguish between the ERO types if they do not bracket or straddle the $4000$\AA~break, which does not work for all EROs (only $z\gtrsim1.4$), and requires high precision photometry \citep[e.g.,][]{stern06}. The mid-IR color techniques introduced by \citet{stern06} ($[3.6]-[24]$) and \citet{wilso07} ($[3.6]-[8.0]$) are more promising because they use the emission from hot dust in star-forming galaxies to obtain a better separation of the two ERO types. However, these mid-IR color techniques require space observatory imaging, which is not always available.  

Accurate spatial clustering measurements require reliable redshift distributions, but there are relatively few EROs with spectroscopic redshifts \citep[e.g.,][]{graha96,dey99,cimat02,conse08}, and spectroscopic samples at high-redshift are vastly incomplete \citep[e.g.,][]{coope07}. Typically, redshift distributions are assumed to have a particular shape or are derived from photometric redshifts or both. This can be a problem because spatial clustering analyses are strongly dependent on the shape of the redshift distribution, contributing to the spread in previous clustering estimates. Unfortunately, using photometric redshifts and models of the redshift distribution remains the best methodology available at present.

An additional problem is that the angular correlation function ($w(\theta)$) is usually fit as a fixed power-law, which may not correctly represent the data even if it is the only option given the sample sizes \citep[e.g.,][]{firth02,georg05}. Furthermore, the uncertainties in the correlation function are frequently assumed to be Poisson with no correlations between data points \citep[e.g.,][]{kong09,kim11}. An incorrect power law, possibly combined with a poor uncertainy model can easily result in incorrect estimates of the spatial correlation length. A likely consequence of these problems is the broad range of ERO clustering estimates that are consistent with their having been the progenitors of anything from local $L^*$ ellipticals to Abell type clusters \citep[see,][for corresponding local galaxy population $r_0$ values, respectively]{norbe02,abadi98}.

In this paper we present a comprehensive measurement of ERO clustering. The sample sizes and survey area are large enough to address all these problems. We consider the largest ERO samples to date, selected in either the $K_s$ or IRAC $[3.6]$ bands over $7.80\rm{~deg}^2$ or $8.53\rm{~deg}^2$ of the NOAO Deep Wide Field Survey (NDWFS) Bo\"otes field \citep{jannu99}, respectively. The NDWFS Bo\"otes field provides a factor of three larger survey area than any previous ERO clustering study. We study three ERO samples using the selection criteria, $(R-K_s)>5.0$, $(I-K_s)>4.0$ and $(I-[3.6])>5.0$, in an attempt to establish the differences and similarities between these samples and to better understand previous results. Our ERO samples contain $28,724$ $(R-K_s)>5.0$, $22,451$ $(I-K_s)>4.0$ and $64,370$ $(I-[3.6])>5.0$ selected EROs, down to magnitude limits of $K_s<19.4$, $K_s<19.4$ and $[3.6]<18.5$, respectively. The sample sizes are 5 times larger than any previous study, and 10 times larger than our previous study of EROs in Bo\"otes \citep{brown05}. 

We use optical NDWFS $B_W RI$ photometry \citep{jannu99}, spectra from the Active Galactic Nuclei (AGN) and Galaxy Evolution Survey \citep[AGES;][]{kocha12} and various small projects with instruments including Keck, Gemini and the Kitt Peak National Observatory Mayall 4m, in conjunction with IRAC photometry from the Spitzer Deep Wide-Field Survey \citep[SDWFS;][]{ashby09}, NEWFIRM $K_s$-band photometry (Gonzalez et al. (in prep.)), and MIPS $24~\mu$m photometry from the MIPS AGN and Galaxy Evolution Survey \citep[MAGES;][]{jannu10}. We test mid-infrared ERO separation techniques utilizing equivalent $B_W RIK_s$ and IRAC photometry \citep[][]{bundy06,dey07,barmb08}, MIPS photometry from the Far-Infrared Deep Extragalactic survey \citep[FIDEL;][]{dicki07} and DEEP2 spectroscopic indicators and 1D spectra \citep[][and references therein]{yan09, newma12} in the Extended Groth Strip (EGS).

This paper is structured as follows. We describe the data sets, photometry and photometric redshift estimation in \S\ref{data}. Our ERO samples are defined and our ERO separation technique is described in \S\ref{Samples}. In \S\ref{analysis} we present our clustering analysis and results, followed by a discussion and summary in \S\ref{discussion}. Throughout this paper, unless otherwise specified, all magnitudes are in the Vega magnitude system. Spitzer IRAC and MIPS magnitudes are quoted using square brackets (e.g., 3.6 $\mu$m is [3.6]). We introduce the following  nomenclature for our ERO samples: $(R-K_s)>5.0$, $(I-K_s)>4.0$ and $(I-[3.6])>5.0$ selected ERO samples are respectively referred to as the ERO$_{RK}$, ERO$_{IK}$ and ERO$_{I3.6}$ samples. We assume a standard $\Lambda$CDM flat cosmology, with $\Omega_m=0.24$, $\Omega_\Lambda=0.76$, $\sigma_8=0.76$, $h=0.73$ and $H_0=100 ~h$ km s$^{-1}$ Mpc$^{-1}$ \citep[][]{sperg07}.

%Data summary - updated 30/10/12 
\section{Data}
\label{data}
\subsection{Object catalogues}
Our galaxies were selected from optical ($B_w RI$) and IR imaging (3.6 to 24.0 $\mu$m) in $8.53$ deg$^2$ of the Bo\"otes field by the NDWFS, SDWFS and MAGES surveys. Where available, we add the NEWFIRM $K_s$-band imaging of $7.80$ deg$^2$ of the Bo\"otes field. Any samples selected using the $K_s$-band are limited to this reduced field size. The wide area AGES survey \citep{kocha12} provides redshifts for the brightest EROs in the field, while the spectra for the fainter EROs come from various small projects using Keck, Gemini and the Kitt Peak National Observatory Mayall $4$m telescope.

Source catalogues were generated by running SExtractor \citep{berti96} in single-image mode on both the $I-$band and $4.5~\mu$m images. Catalogues were merged after accounting for astrometric offsets. We make a pragmatic choice to use the $4.5~\mu$m-selected catalogue for source positions and photometry, as this catalogue was readily available and better represents $3.6~\mu$m sources than $I$-band. This is because objects can be so red as to be detected at $4.5~\mu$m but not in the optical. The colors of $[3.6]<18.5$ galaxies span $0 \lesssim [3.6]-[4.5]  \lesssim 1$, so sources with $[3.6]=18.5$ will have a magnitude of $17.5 \lesssim [4.5] \lesssim 18.5$ and will be comfortably detected in the SDWFS $4.5~\mu$m-selected catalogue since it is 80\% complete at $[4.5]=18.1$ and has a $5\sigma$ depth of $[4.5]\approx18.5$. 

\begin{figure}
\begin{center}
\resizebox{3.3in}{!}{\includegraphics{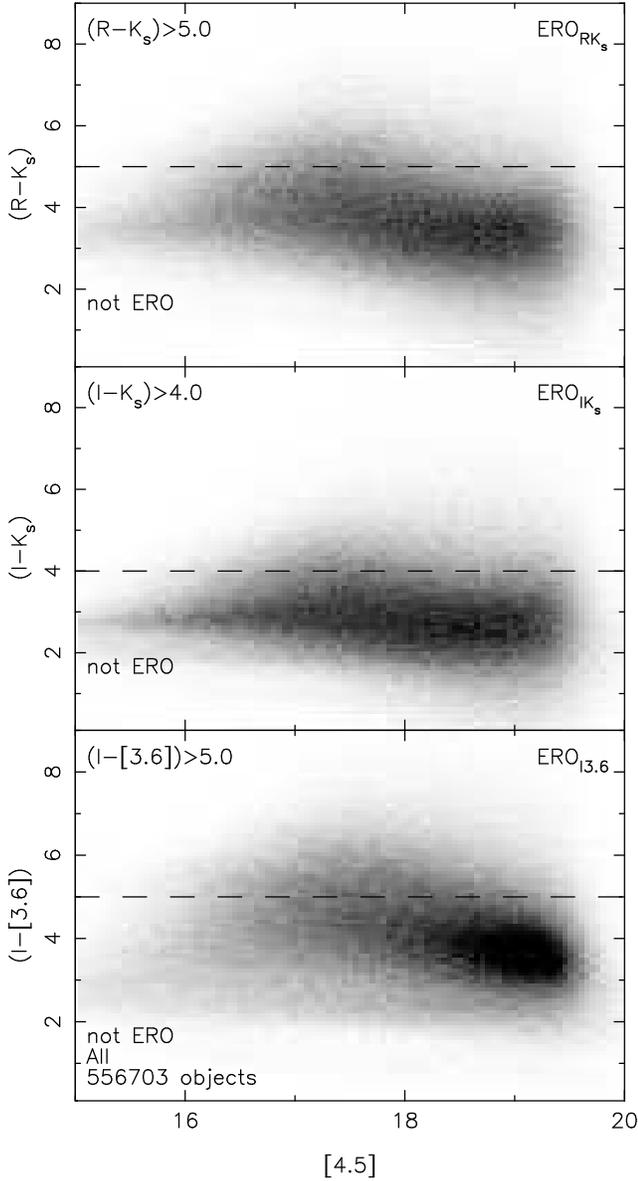}}
\end{center}
\caption{ERO sample selection from the uncleaned $4.5~\mu$m-selected catalogue. Top to bottom, $(R-K_s)$, $(I-K_s)$ and $(I-[3.6])$ color for the entire $[4.5]$-selected catalogue in the NDWFS Bo\"otes field. Our ERO selection criteria $(R-K_s)>5.0$, $(I-K_s)>4.0$ and $(I-[3.6])>5.0$ are shown by the dashed lines.}
\label{sampleselect}
\end{figure}

Photometry in all other bands is measured at the $4.5~\mu$m detected source positions. For the case of objects detected in the $4.5~\mu$m-selected catalogue but with non-detections in $I$-band and $R$-band, we assign conservative upper limits of $I=24$ and $R=25$. These magnitude limits are sufficiently bright so that there are no variations in the galaxy distribution due to the varying depths of the different pointings. We have tested this by inspecting the uniformity of the galaxy distribution for faint magnitude slices about these assigned limits. Our three ERO criteria are shown as applied to the $4.5~\mu$m source catalogue (with imaging quality masks applied to remove unusable areas of the Bo\"otes field) in Figure \ref{sampleselect}. 

We measured aperture photometry using our own analysis code, which estimates uncertainties using a Monte Carlo approach. The optical and near-IR images were smoothed to a common point spread function (PSF) of $1.\arcsec35$ (corresponding to the worst seeing) in the $B_wRIHK$-bands and $1.\arcsec60$ in $J-$band, prior to running the photometry code, so that the fraction of the flux measured with small aperture photometry did not vary with position. The fluxes were PSF corrected and measured using multiple aperture diameters of $4\arcsec-10\arcsec$ in $1\arcsec$ steps, corresponding to physical distances of $\approx24 - 160 ~h^{-1}$kpc at $z=1.0$ respectively. We then assign the default aperture size for color and magnitude measurements using a sliding scale based on the $6\arcsec$ aperture $I-$Band apparent magnitude, where fluxes for objects with $I>23.0$ are measured through a $4\arcsec$ aperture and the fluxes for objects with $I<20$ are measured through $10\arcsec$ apertures, which is sufficient to mitigate aperture bias. Astrometric offsets between bands were measured and corrected. The residuals were far smaller than our aperture size making them negligible. \citet{jannu99} and \citet{brown07} provide additional details for the NDWFS data and \citet{ashby09} provide details for the SDWFS IRAC photometry. 

The $24~\mu$m fluxes were measured using PSF photometry on the MAGES images \citep{jannu10} for each source in the $4.5~\mu$m-selected catalogue, with uncertainties determined using a Monte Carlo method. The NEWFIRM $K_s$ catalogues are described in Gonzalez et al. (in prep.). We used an exposure time-weighted map to exclude poor photometry in the $K_s$-band and we only use photometry from regions where the exposure map was greater than 40\% of the maximum for the NEWFIRM images. Aperture diameters for the NEWFIRM $K_s$ data were determined as described above.

To test some of the techniques used in this paper, we also employ 6,285 galaxies observed through identical $B_wRI$ \citep{dey07}, $K_s$ \citep{bundy06}, IRAC \citep{barmb08}, MIPS \citep[FIDEL;][]{dicki07} filters and with DEEP2 DR3 spectra \citep[][and references therein]{yan09,newma12} from the Extended Groth Strip (EGS). Photometry for these catalogues was measured and matched as described above.

%photometric redshifts  - updated 10/10/12 
\subsection{Photometric Redshifts}
\label{photoz}
\begin{figure}
\begin{center}
\resizebox{3.3in}{!}{\includegraphics{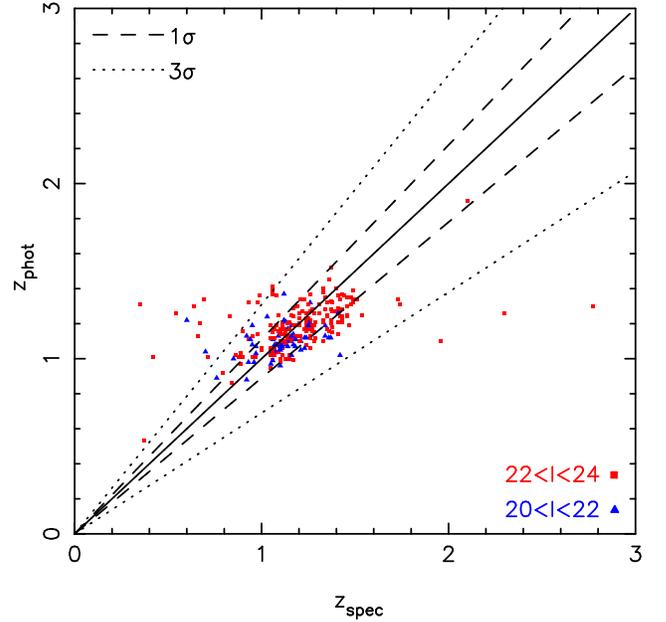}}
\end{center}
\caption{$z_{phot}$ vs. $z_{spec}$ for 273 ERO$_{I3.6}$ galaxies in the NDWFS Bo\"otes field with both photometric ($z_{phot}$) and spectroscopic redshift ($z_{spec}$) estimates. 68\% of our $z_{phot}$ estimates for $20<I<24$ EROs are within $11\%$ (indicated by the dashed lines) of their corresponding spectroscopic redshift ($z_{spec}$), or $\sigma_z/(1+z)\approx0.06$. The results are similar for $20<I<22$ (triangles) and $22<I<24$ (squares), which demonstrate reasonable stability in $z_{phot}$ accuracy between $20<I<24$. The fraction of catastrophic outliers, defined as when $\Delta_z = |z_{phot}-z_{spec}| > 3\sigma_z$, is $\approx6\%$.}
\label{zzplot}
\end{figure}

Photometric redshifts ($z_{phot}$) were estimated using the artificial neural network based ANN$z$ code \citep{firth03}. Our spectroscopic training set was largely derived from AGES \citep[][]{kocha12}, various small NDWFS projects and DEEP2 spectroscopy of the EGS \citep[][]{yan09}. At $z>0.6$ and $I>20.5$, there are relatively few galaxies, so we expanded the training set by making fainter copies of bright galaxies with modified colors (to preserve color-magnitude relations). Objects with colors outside the range of the $z<1.6$ training set were not assigned a $z_{phot}$, as ANN$z$ has a limited ability to extrapolate. Objects with $z_{spec}>1.6$ that are assigned a $z_{phot}$ have systematically underestimated $z_{phot}$ for the same reason, as can be seen in Figure \ref{zzplot}. 

Figure \ref{zzplot} demonstrates the accuracy of our $z_{phot}$ estimation for 273 ERO$_{I3.6}$ galaxies. These galaxies were included in the ANN$z$ training set, which allowed it to span the color of EROs. We find that 68\% of our $z_{phot}$ estimates for $20<I<24$ EROs are within $11\%$ of their corresponding spectroscopic redshift ($z_{spec}$), or $\sigma_z/(1+z)\approx0.06$. The $z_{phot}$ were determined using our $I-$band selected catalogue out to $I<23.5$ and later matched to objects in the $4.5~\mu$m-selected catalogue by astrometry. Due to the limited number of $z_{spec}$ available for our EROs, we cross-checked our $z_{phot}$ against the template $z_{phot}$ of \citet{brodw06} and find good agreement between the two catalogues. 

The fraction of catastrophic $z_{phot}$ errors, which we define as when when $\Delta_z = |z_{phot}-z_{spec}| > 3\sigma_z$, is $\approx6\%$. At $z\approx0.5$, stellar spectral features around $\approx2.3\mu$m produce red $([3.6]-[4.5])$ colors. This can confuse ANN$z$ and is likely responsible for the systematic overestimation of $z_{phot}$ for objects with $0.4<z_{spec}<0.7$ shown in Figure \ref{zzplot}. The likely reason for systematic $z_{phot}$ underestimates of objects with $z_{spec}>1.6$ is as described earlier. 

We note that our spectra do not uniformly cover the magnitude and redshift range of our sample. There are very few $z>1.5$ galaxy spectra in Bo\"otes and a large number of $z<0.8$ spectra from AGES. Since the very red nature of EROs limits the redshift of our sample to $z\gtrsim0.8$, the fraction of $z\approx0.5$ galaxies contaminating our sample is probably overestimated in Figure \ref{zzplot}. We discuss our method for managing these $z_{phot}$ errors, as well as our method for estimating the overall redshift distributions for our samples in \S\ref{dndz}.

%catalogue cleaning  - updated 27/11/12 
\subsection{Star and Quasar removal}
\label{sampleclean}
\begin{figure*}
\begin{center}
\resizebox{4.1in}{!}{\includegraphics{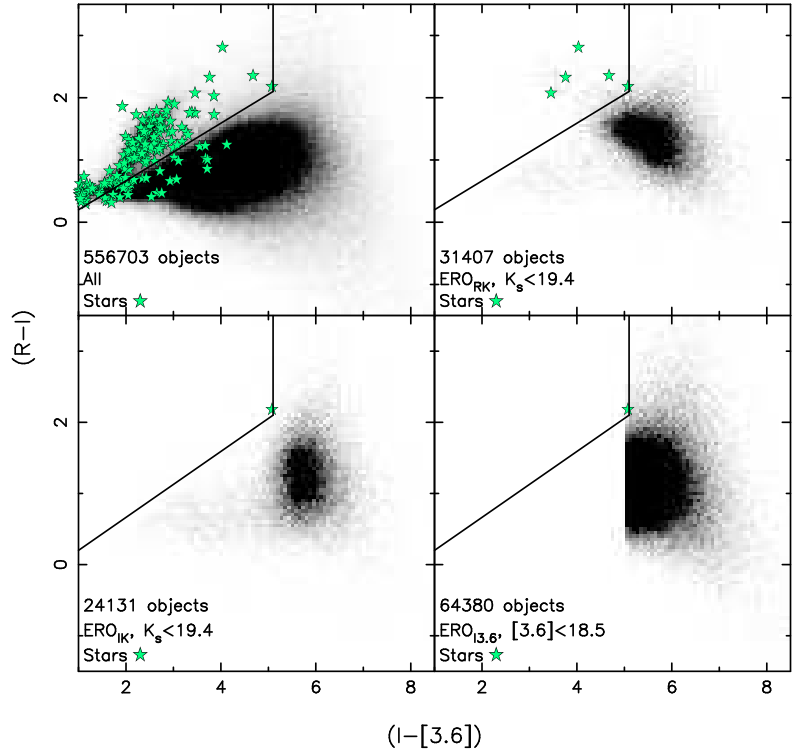}}
\end{center}
\caption{Control of stellar contamination shown for the uncleaned $4.5~\mu$m-selected catalogue. $(R-I)$ vs. $(I-[3.6])$ for All, ERO$_{RK}$, ERO$_{IK}$ and ERO$_{I3.6}$ objects in the NDWFS Bo\"otes field. Green stars are AGES spectroscopically identified stars. We use the location of these stars to place an empirical cut to remove stars from our ERO samples. The stellar cut combined with the intrinsically red ($I-[3.6]$) color of EROs keeps the samples free from stellar contamination.}
\label{STARremove}
\end{figure*}

AGES spectra were used to identify 241 stars and 464 quasars (QSOs) in our $4.5~\mu$m-selected catalogue. Some of the AGES selected stars will have peculiar colors because they were targeted as candidate AGN (e.g., X-ray sources). We used the ($R-I$) and ($I-[3.6]$) color distribution of these stars to define a cut to remove these from our ERO samples. This star removal cut, of sources with $(R-I)>0.46(I-[3.6])-0.26,~\rm{and~with}~(I-[3.6])<5.1$, which is shown in Figure \ref{STARremove}, removes $58,790$ objects from the $4.5~\mu$m-selected catalogue. Figure \ref{STARremove} demonstrates that due to the extremely red color of EROs, this star cut excludes 100\% of stars with a negligible impact on the ERO samples.

\begin{figure*}
\begin{center}
\resizebox{4.1in}{!}{\includegraphics{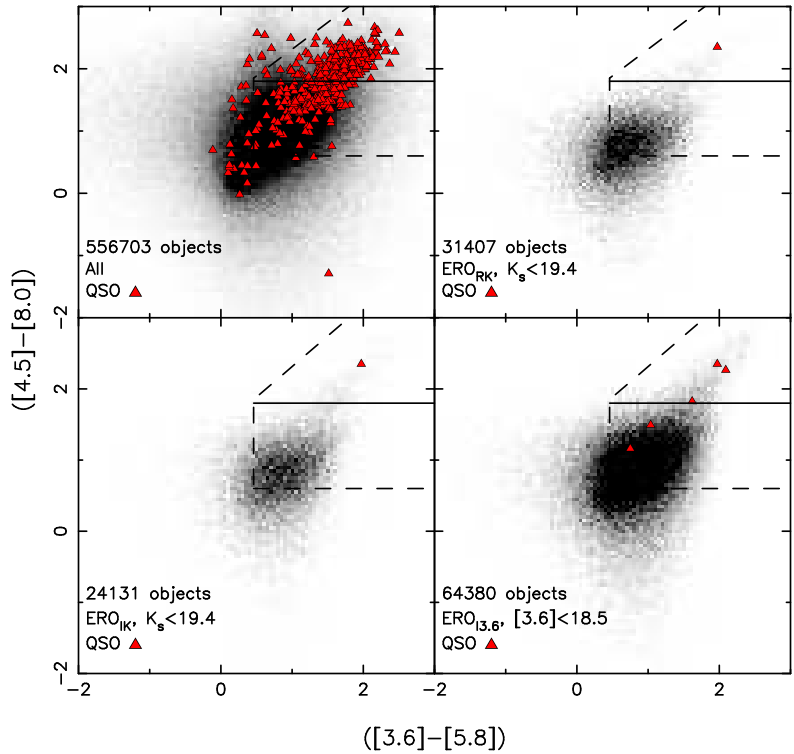}}
\end{center}
\caption{Control of quasar contamination shown for the uncleaned $4.5~\mu$m-selected catalogue. A modified form of the \citet{lacy04} mid-IR, ($[4.5]-[8.0]$) vs. ($[3.6]-[5.8]$) AGN selection region is used to minimise quasar contamination within our sample. This is shown for All, ERO$_{RK}$, ERO$_{IK}$ and ERO$_{I3.6}$ objects in the NDWFS Bo\"otes field. AGES spectroscopically identified QSOs are shown with red triangles and we use these to place a more conservative empirical cut at $([4.5]-[8.0])\geqslant1.8$ (solid line), to remove obvious AGN contaminants from our sample. The original \citet{lacy04} criteria are shown by the dashed lines.}
\label{AGNremove}
\end{figure*}

There are multiple IRAC color AGN selection criteria in the literature \citep[e.g.,][]{lacy04,stern05}. Due to the ($[3.6]-[4.5]$) colors of $z\gtrsim1$ galaxies, the \citet{stern06} AGN criterion deliberately removes a large number of EROs. Using the colors of AGES identified quasars, we demonstrate in Figure \ref{AGNremove} the effectiveness of the IRAC color selection introduced by \citet{lacy04} for the ERO samples. Figure \ref{AGNremove} shows that quasars selected by our ERO criteria occupy a region of color space corresponding to a distinct tail within the \citet{lacy04} AGN region. We have modified the \citet{lacy04} color criteria to conservatively select AGN in this distinct tail seen above the ERO populations in Figure \ref{AGNremove}. It is clear that this selection criteria is not 100\% effective, but our aim is to select a complete ERO population, not a complete sample of AGN. The AGN cut removes $71,664$ AGN from the $4.5~\mu$m-selected catalogue.

After application of both of these stellar and QSO removal cuts, our $4.5~\mu$m source catalogue is reduced from $556,703$ to $426,249$ objects. Since the $K_s$-band photometry was filtered by an exposure time weight map, the $4.5~\mu$m source catalogue including good $K_s$-band photometry is reduced to $366,070$ objects.    

%ERO sample selection and comparison - updated 27/11/12 
\section{Samples}
\label{Samples}
\subsection{ERO selection}
\label{samples}
\begin{figure}
\begin{center}
\resizebox{3.3in}{!}{\includegraphics{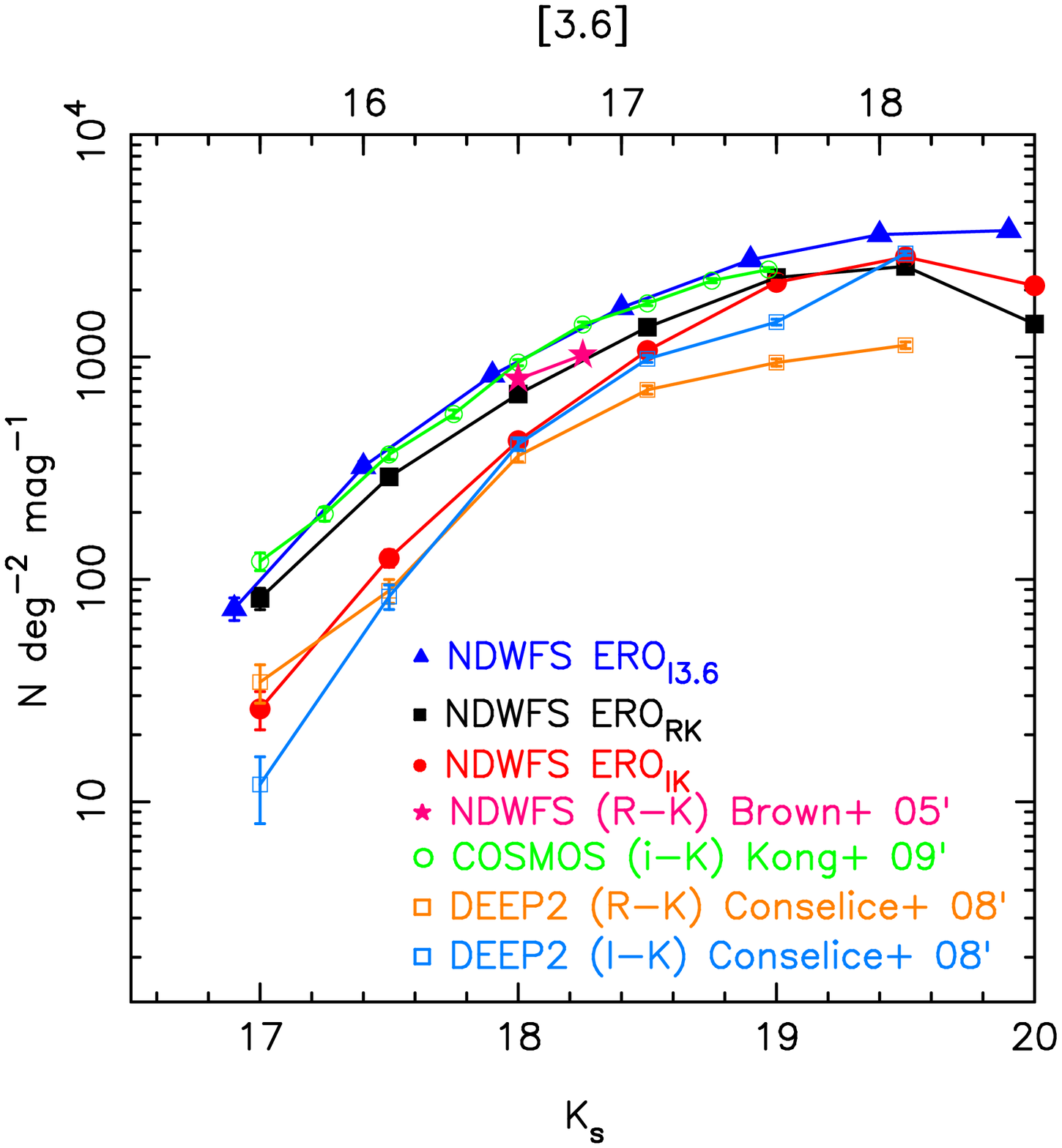}}
\end{center}
\caption{Completeness-corrected surface densities ($N$) for each ERO sample. For comparison with the surface densities found in previous ERO studies we show the distributions from \citet{kong09} ($(i-K)>3.96$), \citet{brown05} ($(R-K)>5.0$) and \citet{conse08} ($(R-K)>5.3$) and $((I-K)>4.0)$. The error bars are included for all points, but in many cases are too small to be seen. We use $K_s\approx[3.6]+1.4$ to shift $[3.6]$ magnitudes onto the $K_s$ scale, based on the average of $K_s-[3.6]$ for our EROs.}
\label{nden}
\end{figure}

From our cleaned $4.5~\mu$m source catalogues, the ERO$_{RK}$, ERO$_{IK}$ and ERO$_{I3.6}$ cuts respectively select 16\%, 17\% and 27\% of all $4.5~\mu$m detected sources. From $60,066$, $63,798$ and $113,098$ $4.5~\mu$m-selected ERO$_{RK}$, ERO$_{IK}$ and ERO$_{I3.6}$ galaxies, we selected $28,724$, $22,451$ and $64,370$ EROs with $K_s<19.4$, $K_s<19.4$ and $[3.6]<18.5$, respectively. Sample magnitude limits were set to correspond to the flattening in surface density ($N$) seen in Figure \ref{nden}. SDWFS is $80\%$ complete at  $[3.6]=18.2$, and this limit is also consistent with our $5\sigma$ $I-$band and $[3.6]$ detection limits \citep{jannu99,ashby09}. Throughout this paper we take $K_s\approx[3.6]+1.4$, based on the average of $K_s-[3.6]$ for our EROs. 

The fraction of $4.5~\mu$m-selected sources meeting the ERO$_{I3.6}$ criterion was found to increase with increasing limiting magnitude, from $19\%$ at $[3.6]<17.5$ to $57\%$ at $[3.6]<18.5$. Similarly, the respective fractions of $4.5~\mu$m-selected sources meeting the ERO$_{RK}$ and ERO$_{IK}$ criteria were also found to increase with increasing limiting magnitude from $11\%$ and $6\%$ at $K_s<18.4$ to $48\%$ and $35\%$ at $K_s<19.4$. The fractional increase in EROs with increasing apparent magnitude for the ERO$_{RK}$ and ERO$_{IK}$ samples is $29-37\%$. The ERO$_{I3.6}$ sample has a fractional increase of $38\%$ and selects a much larger number of high-redshift galaxies than both ERO$_{RK}$ and ERO$_{IK}$ samples. The ERO$_{RK}$ criterion selects a greater fraction than the ERO$_{IK}$ criterion, this is likely due to the difference in mean redshift between these two samples. 

As shown in Figure \ref{nden}, the surface density of the ERO$_{RK}$ sample is consistent with that of our previous study of a sub-field of Bo\"otes for the same selection criteria \citep{brown05}. Furthermore, if we raise our criteria to $(R-K_s)>5.3$ to match that of \citet{conse08} we find good agreement in surface density. Likewise, we see good agreement between our ERO$_{IK}$ sample and the $(I-K)>4.0$ EROs of \citet{conse08}. To compare with the $(i-K)>3.96$ EROs of \citet{kong09}, we tested the surface density of $(I-K_s)>3.7$ (which roughly corresponds to $(i-K)>3.96$) objects and also found good agreement in surface densities.

\begin{figure*}
\begin{center}
\includegraphics[scale=0.48]{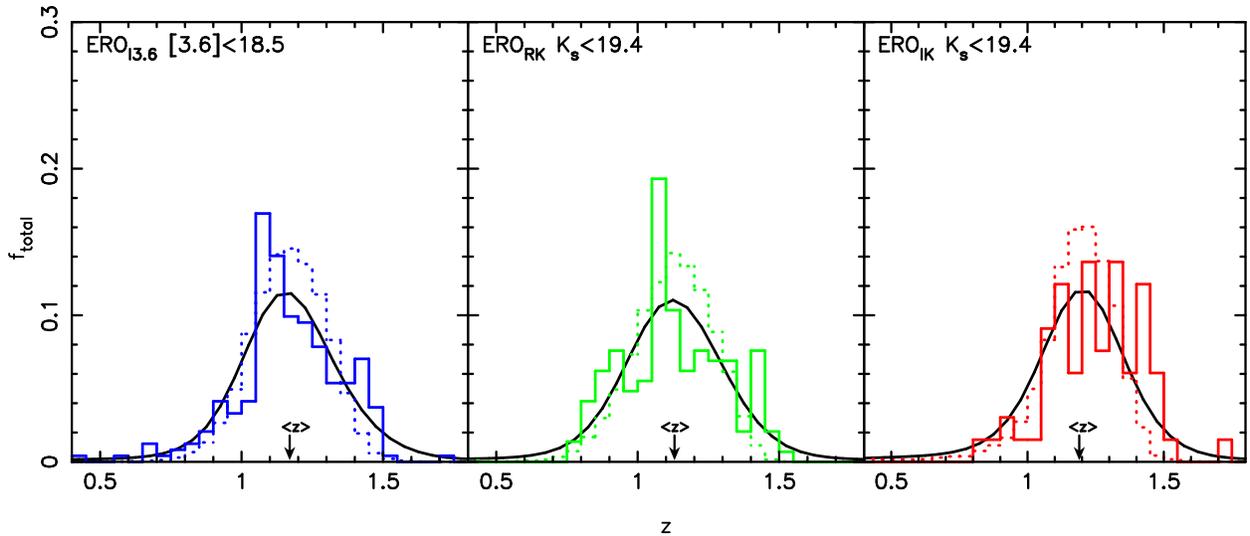}
\end{center}
\caption{Normalised photometric (dashed histogram) and spectroscopic (solid histogram) redshift distributions as compared to our model distribution (solid line) for (from left to right) ERO$_{I3.6}$ (blue lines) with $[3.6]<18.5$, ERO$_{RK}$ (green lines) with $K_s<19.4$, and ERO$_{IK}$ (red lines) with $K_s<19.4$ samples. We find that the redder selection criteria (ERO$_{I3.6}$ and ERO$_{IK}$) select samples at higher $\bar{z}$, as shown by the $\langle z\rangle$ marker in each panel. While the redshift bins here have $\Delta z=0.05$ to reduce the noise for the spectroscopic redshifts, $\Delta z=0.02$ is used in our analysis.}
\label{dndzcompare}
\end{figure*}

In Figure \ref{dndzcompare} we show the model redshift distributions for EROs out to the magnitude limits for each of our samples. The mean redshift ($\bar{z}$) of the ERO$_{RK}$ sample, $\bar{z}=1.13$, is lower than that of the ERO$_{IK}$ sample, with $\bar{z}=1.19$, in agreement with previous work \citep[e.g.,][]{conse08,kim11}. We find that the ERO$_{I3.6}$ sample has a very similar redshift distribution to the ERO$_{IK}$ sample, with $\bar{z}=1.17$, although with far greater number density. Since all our samples contain at least $\gtrsim9000$ objects with photometric redshifts, the statistical uncertainties in the mean redshift are negligible ($<1\%$) and we are dominated by systematic uncertainties.

Figures \ref{STARremove} and \ref{AGNremove} illustrate in color space the similarities and differences between the ERO$_{I3.6}$, ERO$_{RK}$ and ERO$_{IK}$ samples. We see that there is significant overlap between the samples selected by the three ERO selection criteria. Quantitatively, the bulk ($>86\%$) of the ERO$_{IK}$ sample is a sub-sample of both the ERO$_{I3.6}$ and ERO$_{RK}$ samples. Of the ERO$_{RK}$ sample, $78\%$ have $(I-[3.6])>5.0$, however only $\approx55\%$ of the ERO$_{I3.6}$ sample have $(I-K_s)>4.0$ and $(R-K_s)>5.0$. The differences are likely due to the ERO$_{I3.6}$ having a greater surface density because at $3.6~\mu$m we are detecting rest frame NIR emission which is less obscured by dust. Thus, both the ERO$_{RK}$ and ERO$_{IK}$ samples are largely sub-sets of the ERO$_{I3.6}$ sample, but with different mean redshift. To determine whether there is a relationship between ERO clustering strength and luminosity, we separate each ERO sample into four $[3.6]$ and $K_s$ magnitude-limited sub-samples as defined in Tables \ref{resultssum1}  and \ref{resultssum3}. Again, since all of our samples contain at least $\gtrsim1000$ objects with photometric redshifts, the statistical uncertainties in the mean redshifts are negligable.

%separation of ERO mixed population - updated 27/11/12 
\subsection{ERO Type Separation}
\label{separation}
At low redshift the clustering of passive galaxies differs from that of star-forming galaxies \citep[e.g.,][]{zehav10}, but this may change with redshift \citep[e.g., $z\gtrsim1.3 - 2.0$; ][]{coope07,brodw08}. There is no reason to expect passive and star-forming galaxies to occupy the same halos and have identical clustering properties. It is thus both interesting and necessary to separate star-forming EROs from passive EROs in order to determine if they show differing clustering properties. If star-forming EROs have distinctly different clustering properties from their passive counterparts, it would show that they are two distinct population of galaxies, and it would mean that they are the progenitors of two very different populations of local Universe galaxies. It would also mean that clustering studies of mixed ERO populations cannot be used to accurately link EROs to local galaxy populations.

To independently measure the clustering of passive and star-forming EROs, we use the wealth of data in the EGS to explore mid-IR color separation techniques using IRAC \citep{wilso07} and MIPS \citep{stern06} colors. These techniques are physically motivated by polycyclic aromatic hydrocarbon (PAH) emission from warm dust in star-forming regions. The $([3.6]-[8.0])$ color is able to separate the ERO types as long as the strong PAH emission features lie in $8.0~\mu$m band, but this ceases to be true for $z\gtrsim1$. \citet{stern06} showed that the difference in $([3.6]-[8.0])$ color between a star-forming ERO and a passive ERO can be very small for archetypal star-forming and passive EROs. \citet{kong09} also find that very high precision photometry is required to produce good agreement between $([3.6]-[8.0])$ color and morphological ERO classifications. We tested the $([3.6]-[8.0])$ color of EROs in the EGS using DEEP2 spectroscopic indicators. Although we do observe a tendency for passive EROs to have a bluer $([3.6]-[8.0])$ color than star-forming ones, the two populations are not well separated and this technique is not suitable for this work.

\begin{figure}
\begin{center}
\includegraphics[scale=0.46]{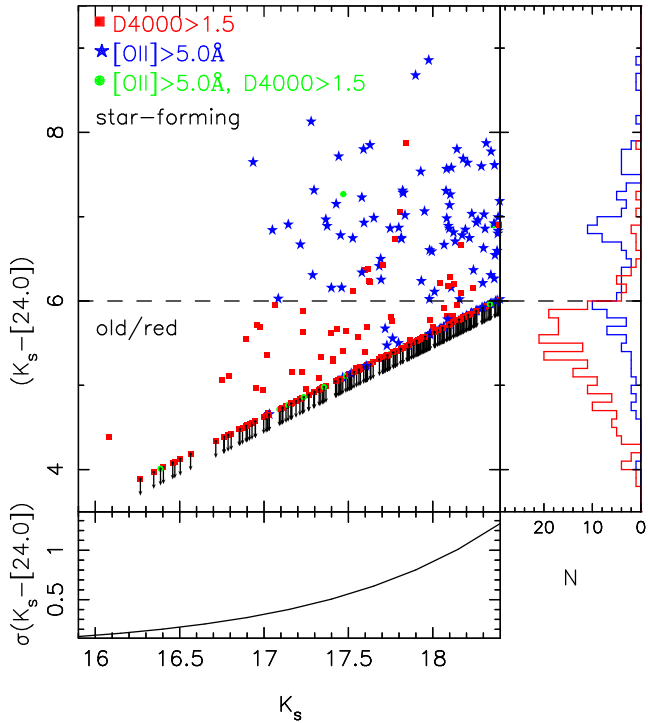}
\end{center}
\caption{Galaxy type separation using $(K_s-[24])$ colors as a function of $K_s$ for all $0.6<z<1.5$ galaxies in the EGS. Objects with $f_{24}<80~\mu$Jy are indicated by arrows. Sources are marked by whether they have D4000$>1.5$ indicating an old stellar population (squares), [OII]$>5.0\AA$ indicating star formation (stars) or both (circles). We empirically separate the populations at $(K_s-[24])=6.0$. The FIDEL fluxes have been degraded in precision to match the noise and detection limit of the Bo\"otes data. The histogram in the side panel shows the color distribution of D4000$>1.5$ sources (red line) and [OII]$>5.0\AA$ sources (blue line). The lower panel shows $\sigma(K_s-[24])$ as a function of apparent magnitude based on a $24~\mu$m RMS of $40~\mu\rm{Jy}$.}
\label{EROsepEGS_Ks}
\end{figure}

\citet{stern06} showed that the $(K_s-[24])$ color has a much broader color separation than the $([3.6]-[8.0])$ color for archetypal star-forming and passive EROs. In Figure \ref{EROsepEGS_Ks}, $(K_s-[24])$ color versus $K_s$ magnitude is shown for $0.6<z<1.5$ galaxies in the EGS that have $B_wRIK_s$, IRAC, $24~\mu$m and DEEP2 detections. We independently identify passive EROs using the DEEP2 D4000 spectroscopic indicator, which provides a measure of the strength of the $4000$\AA{} break. D4000 $>1.5$ is used to indicate an evolved stellar population. To independently identify the star-forming EROs, we use a DEEP2 rest-frame [OII] equivalent width $>5.0$\AA~to identify galaxies with significant active star formation. Only DEEP2 spectroscopic indicators with a $\geqslant3\sigma$ detection were used in this analysis. 

The FIDEL $24~\mu$m catalogue has a $3\sigma$ depth of $30~\mu$Jy, so we use this as the upper limit on any source without a $24~\mu$m detection. The Bo\"otes field $24~\mu$m data (MAGES) has an RMS of $40~\mu$Jy and thus a $2\sigma$ depth of $80~\mu$Jy. To determine how the noisier and shallower MAGES data will effect population separation criteria, in Figure \ref{EROsepEGS_Ks} we have applied additional Gaussian scatter to the FIDEL data to mimic the RMS noise level of MAGES. The two populations start to merge at $K_s=18.4$ and $[3.6]=17.5$,  so the $2\sigma$ detection limit of MAGES prevents us separating the two populations at fainter magnitude.

\begin{figure}
\begin{center}
\includegraphics[scale=0.46]{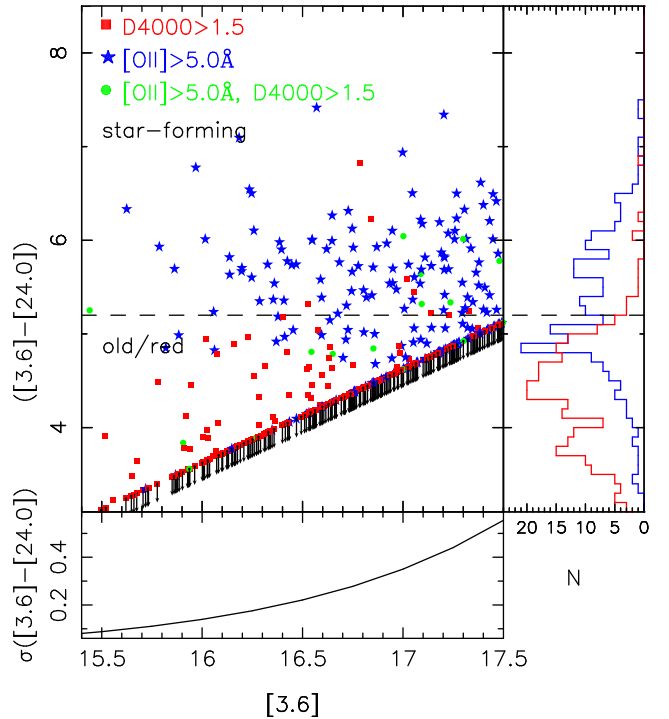}
\end{center}
\caption{Galaxy type separation using $([3.6]-[24])$ colors as a function of $[3.6]$ for all $0.6<z<1.5$ galaxies in the EGS. Objects with $f_{24}<80~\mu$Jy are indicated by arrows. Sources are marked by whether they have D4000$>1.5$ indicating an old stellar population (squares), [OII]$>5.0\AA$ indicating star formation (stars) or both (circles). We empirically separate the populations at $([3.6]-[24])=5.2$. The FIDEL fluxes have been degraded in precision to match the noise and detection limit of the Bo\"otes data. The histogram in the side panel shows the color distribution of D4000$>1.5$ sources (red line) and [OII]$>5.0\AA$ sources (blue line). The lower panel shows $\sigma([3.6]-[24])$ as a function of apparent magnitude based on a $24~\mu$m RMS of $40~\mu\rm{Jy}$.}
\label{EROsepEGS_CH1}
\end{figure}

Since the $3.6~\mu$m imaging in SDWFS is significantly deeper than the NEWFIRM $K_s$ in Bo\"otes, not every source with a $3.6~\mu$m detection will have a good $K_s$ detection. To achieve better separation for the ERO$_{I3.6}$ sample we tested using the $([3.6]-[24])$ color as a separation method. Figure \ref{EROsepEGS_CH1} shows that this also works in a similar manner to the $(K_s-[24])$ color. In both Figures \ref{EROsepEGS_Ks} and \ref{EROsepEGS_CH1} the EROs with strong D4000 and those with strong [OII] are well separated. This is further supported by the histograms in Figures \ref{EROsepEGS_Ks} and \ref{EROsepEGS_CH1} which show a bimodal distribution in color, but with some overlap, for these two spectral indicators. Some galaxies have both strong D4000 and [OII]. For these objects we visually inspected the DEEP2 1D spectra to confirm the color classification and found that it accurately distinguished between a passive and star-forming galaxies in at least 90\% of cases. 

We empirically place a discriminator between passive and star-forming EROs where the base of the strong D4000 distribution (seen in Figures \ref{EROsepEGS_Ks} and \ref{EROsepEGS_CH1}) and the respective color for an object at the $24~\mu$m detection limit meet. Using a $2\sigma$ detection limit for a MAGES source, we can use this separation technique out to $K_s=18.4$ and $[3.6]=17.5$, with discriminators at $(K_s-[24])=6.0$ and $([3.6]-[24])=5.2$, respectively. We consider $2\sigma$ a significant limit for the purposes of this work as we are interested in whether the source is emitting at $24~\mu$m rather than its specific flux. Since the placement of these discriminators is empirical, we varied the discriminators by $\pm0.2$, and found no significant change in the separated samples and final clustering results.   

We define objects with $(K_s-[24])>6.0$ and $([3.6]-[24])>5.2$ as star-forming	 and those with $(K_s-[24])\leqslant6.0$ and $([3.6]-[24])\leqslant5.2$ as passive for $K_s<18.4$ and $[3.6]<17.5$, respectively. Details for the separated ERO$_{RK}$, ERO$_{IK}$ and ERO$_{I3.6}$ samples are given in Tables \ref{resultssum2} and \ref{resultssum4}, including the sample size, magnitude limit, $\bar{z}$ and relative fractions of the combined ERO samples to the same limiting magnitude. As a check, the archetypal star-forming ERO HR10 \citep[][]{graha96} with $(K_s-[24])=7.61$ and $([3.6]-[24])=5.82$ \citep{stern06} is clearly classified as star-forming, while the archetypal passive galaxy LBDS 53W091 \citep{stern06} with $(K_s-[24])<5.56$ and $([3.6]-[24])<4.13$ (assuming a $24~\mu$m flux upper limit of $40~\mu$Jy) is clearly classified as passive.

There is some contamination on both sides of our color criteria and the level of contamination increases for fainter sources in both Figures \ref{EROsepEGS_Ks} and \ref{EROsepEGS_CH1}. One possible physical reason for this is that PAH emission at observed $24~\mu$m varies with redshift. There is also rest-frame $10~\mu$m silicate absorption which would be observed at $24~\mu$m from a $z\approx1.3$ ERO. Contamination of this nature would therefore increase for fainter EROs since these are more likely to have higher redshifts. 

Alternatively, EROs are probably not simply archetypal passive and star-forming galaxies. If we assume that the SED of a passive galaxy with little dust roughly follows a blackbody at long wavelengths, we can approximate the expected $24~\mu$m  flux by the Rayleigh-Jeans tail. For LBDS 53W091 with a $3.6~\mu$m flux of $35~\mu \rm{Jy}$, we expect an approximate $24~\mu$m flux of $0.02~\mu\rm{Jy}$ and so should have no $24~\mu$m detections for truly passive galaxies. However, there are objects with both a $24~\mu$m detection and strong D4000 in both Figures \ref{EROsepEGS_Ks} and \ref{EROsepEGS_CH1}. The $(K_s-[24])$ and $([3.6]-[24])$ colors are effectively sampling the specific star formation rate (SSFR) of these galaxies. Therefore, these mixed cases are likely passive galaxies with some star formation and/or an AGN. To cross-check that this is the case, we further examine our cut against purely $24~\mu$m non-detected and detected samples, indicating star formation rate (SFR) only (discussed in \S\ref{spatial}). 

\begin{figure}
\begin{center}
\includegraphics[scale=0.46]{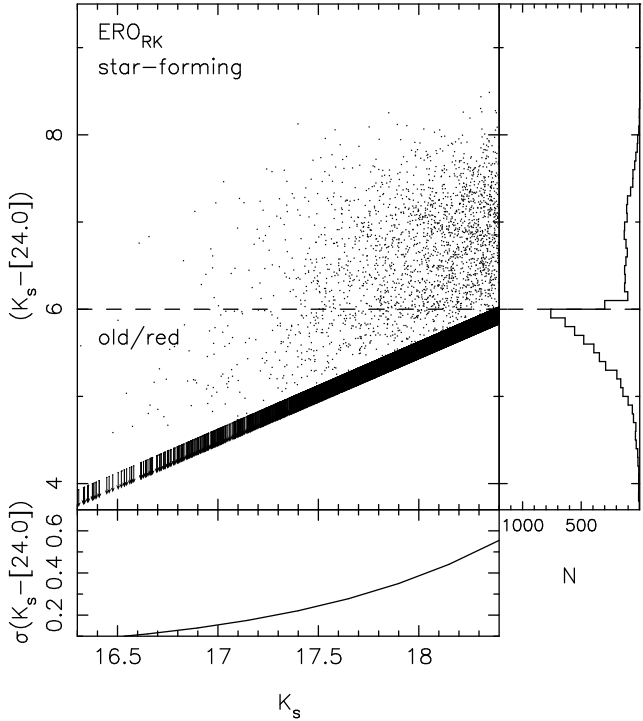}
\end{center}
\caption{Separating star-forming and passive EROs using the $(K_s-[24])$ color in the ERO$_{RK}$ sample. We divide the populations at the dashed line $(K_s-[24])=6.0$. Upper limits $f_{24lim}$ are marked with arrows. The histogram in the side panel shows the relative distribution in color space for Bo\"otes objects down to $K_s<18.4$. The lower panel shows $\sigma(K_s-[24])$ as a function of apparent magnitude based on a $24~\mu$m RMS of $40~\mu \rm{Jy}$.}
\label{EROsepBootes_Ks}
\end{figure}

\begin{figure}
\begin{center}
\includegraphics[scale=0.46]{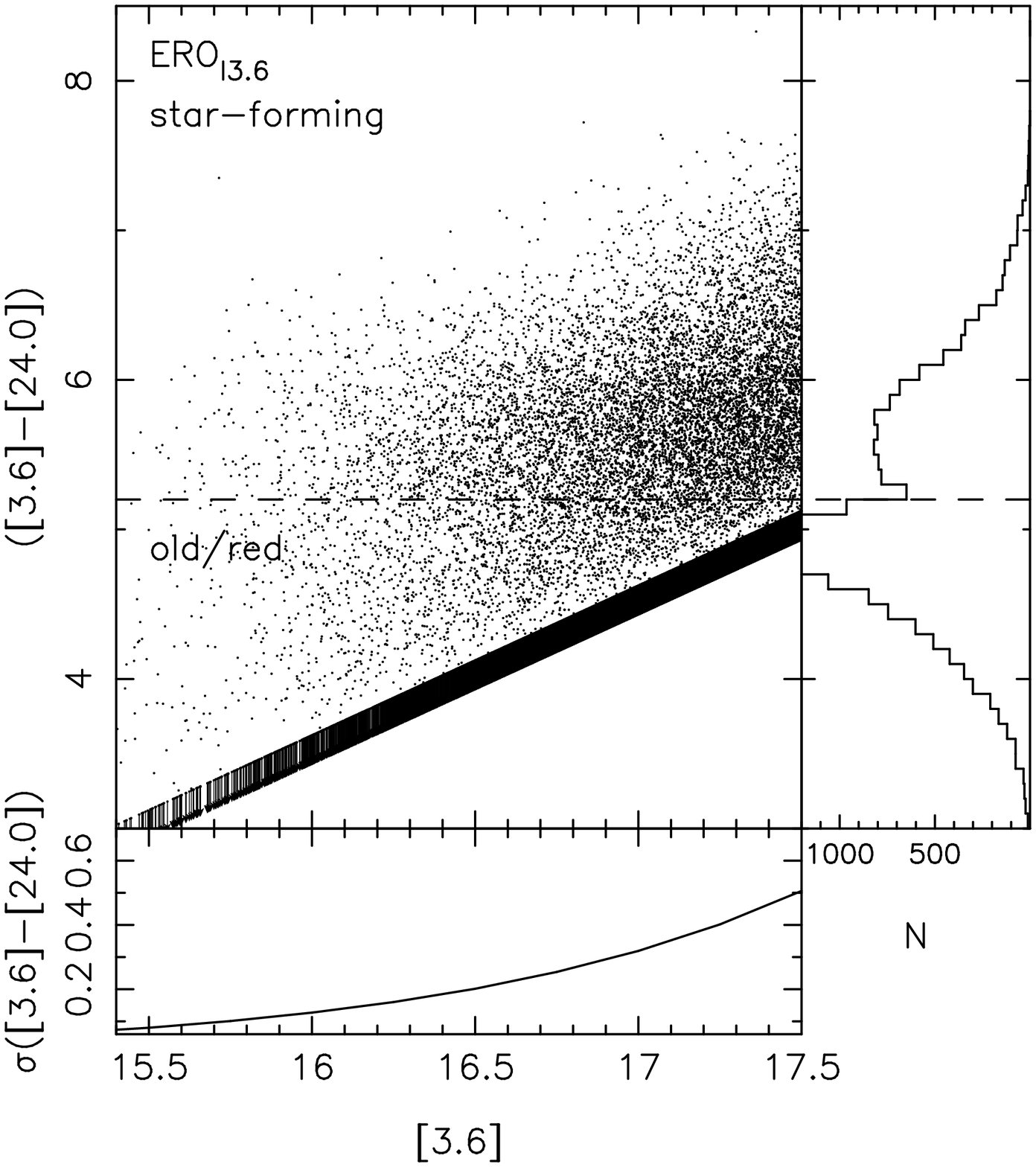}
\end{center}
\caption{Separating star-forming and passive EROs using the $([3.6]-[24])$ color in the ERO$_{I3.6}$ sample. We divide the populations at the dashed line $([3.6]-[24])=5.2$. Upper limits $f_{24lim}$ are marked with arrows. The histogram in the side panel shows the relative distribution in color space for Bo\"otes objects down to $[3.6]<17.5$. The lower panel shows $\sigma([3.6]-[24])$ as a function of apparent magnitude based on a $24~\mu$m RMS of $40~\mu \rm{Jy}$.}
\label{EROsepBootes_CH1}
\end{figure}

We show the application of these separation criteria to the Bo\"otes field ERO$_{RK}$ and ERO$_{I3.6}$ samples in Figures \ref{EROsepBootes_Ks} and \ref{EROsepBootes_CH1}, respectively. The star-forming EROs form a locus that broadens toward fainter magnitudes. The broadening is largely an effect of the growth in the $(K_s-[24])$ and $([3.6]-[24])$ uncertainties with magnitude, which we show in the lower panel of each Figure. Despite the broadening, it is clear that the $(K_s-[24])$ and $([3.6]-[24])$ colors are finding distinct populations. This is clearly evident in the bimodal distribution in ($[3.6]-[24]$) color shown by the histogram in Figure \ref{EROsepBootes_CH1}. Figure \ref{EROsepBootes_Ks} shows weaker evidence for this bi-modality in the ($K_s-[24]$) color, which may be due to the reduced fraction of star-forming EROs in the ERO$_{RK}$ sample (see, Tables \ref{resultssum2} and \ref{resultssum4}).

The relative fractions of star-forming and passive EROs varies between the different ERO selection criteria. At $K_s<18.4$, the ERO$_{RK}$ sample shows a star-forming fraction of 30\%. Similarly, at $K_s<18.4$, the ERO$_{IK}$ sample shows a star-forming fraction of 32\%. In contrast, at $[3.6]<17.5$, the ERO$_{I3.6}$ sample has a significantly higher star-forming fraction of 41\%. The star-forming fractions found for the ERO$_{RK}$ and ERO$_{IK}$ samples are consistent with previous studies \citep[e.g., the fraction of peculiar galaxies;][$(R-K_s)>5.3$]{conse08}. The relative fraction of star-forming EROs in the ERO$_{I3.6}$ sample is consistent with that found by \citet{kong09} using $(i-K)>3.96$. \citet{moust04} used a $(R-K_s)>5.0$ ERO cut and found a passive ERO fraction of only $33-44\%$, but this study went to a much fainter $K_s<22$ flux limit. We observe in all three ERO samples an increase of star-forming ERO fraction with increasing apparent magnitude (see Tables \ref{resultssum2} and \ref{resultssum4}, and Figures \ref{EROsepEGS_Ks}, \ref{EROsepEGS_CH1}, \ref{EROsepBootes_Ks}, and \ref{EROsepBootes_CH1}) and this is  consistent with the fraction of star-forming EROs increasing with decreasing luminosity and stellar mass \citep[e.g.,][]{moust04,conse08,kong09,kim11}.

%Analysis - updated 27/11/12 
\section{Analysis}
\label{analysis}
%Analysis introduction
In this section we analyze the angular and spatial clustering of our ERO samples. Using these methods, we place the EROs in an evolutionary sequence and make links to local galaxy populations. In \S\ref{ACF} we study the angular clustering of each ERO sample as a function of limiting magnitude and ERO color and selection criterion. The three ERO samples studied, ERO$_{RK}$, ERO$_{IK}$ and ERO$_{I3.6}$, allow us to explore this to some degree across different redshift intervals. We build model redshift distributions in \S\ref{dndz} and discuss the effects and limitations of the photometric redshifts used to generate them. We combine the angular clustering and the redshift distribution models to estimate the spatial correlation function in \S\ref{spatial}. We also analyze the spatial correlation function as a function of limiting magnitude and ERO color and selection criterion. We use the clustering of EROs and dark matter halos to connect EROs to other galaxy populations across redshift in \S\ref{link}. Finally we estimate the host dark matter halo masses of our EROs in \S\ref{halomass}. 

\subsection{The angular correlation function}
\label{ACF}
We measured the angular correlation function for each magnitude limited sample using the \citet{landy93} estimator,
\begin{equation}
\label{landy}
\hat{w}(\theta)=\frac{DD-2DR+RR}{RR},
\end{equation}
where, DD, DR and RR are the number of Data-Data, Data-Random and Random-Random galaxy pairs at angular separation $\theta\pm\delta\theta/2$ respectively. Pairs were counted in logarithmically spaced angular bins with $\Delta \rm{log_{10}}{~\theta}=0.25$. Since any survey area represents a subset of the Universe, estimators such as Equation(\ref{landy}) are subject to an integral constraint \citep{groth77},
\begin{equation}
\label{intcon}
\int\int\hat{w}(\theta_{12})d\Omega_1d\Omega_2\simeq0
\end{equation}
where $\theta_{12}$ is the angular separation between solid angle elements $d\Omega_1$ and $d\Omega_2$. Since Equation(\ref{landy}) is normalized by itself, the integral constraint leads to a systematic underestimate of $\hat{w}(\theta)$ which is corrected by the addition of the variance term,
\begin{equation}
\label{variance}
\sigma^2=\frac{1}{\Omega^2}\int{\int{\hat{w}(\theta_{12})d\Omega_1}d\Omega_2}.
\end{equation}
This corrects for the contribution to clustering from the variance in number counts over the finite survey area $\Omega$. We estimate the uncertainty in each angular correlation function bin using a modification to the Gaussian approximation to the covariance matrix discussed in \citet{brown08}. Our measured $w(1')$ values are well below unity, so we consider our uncertainty estimates valid to first order.    

\begin{figure*}
\begin{center}
\includegraphics[scale=0.42]{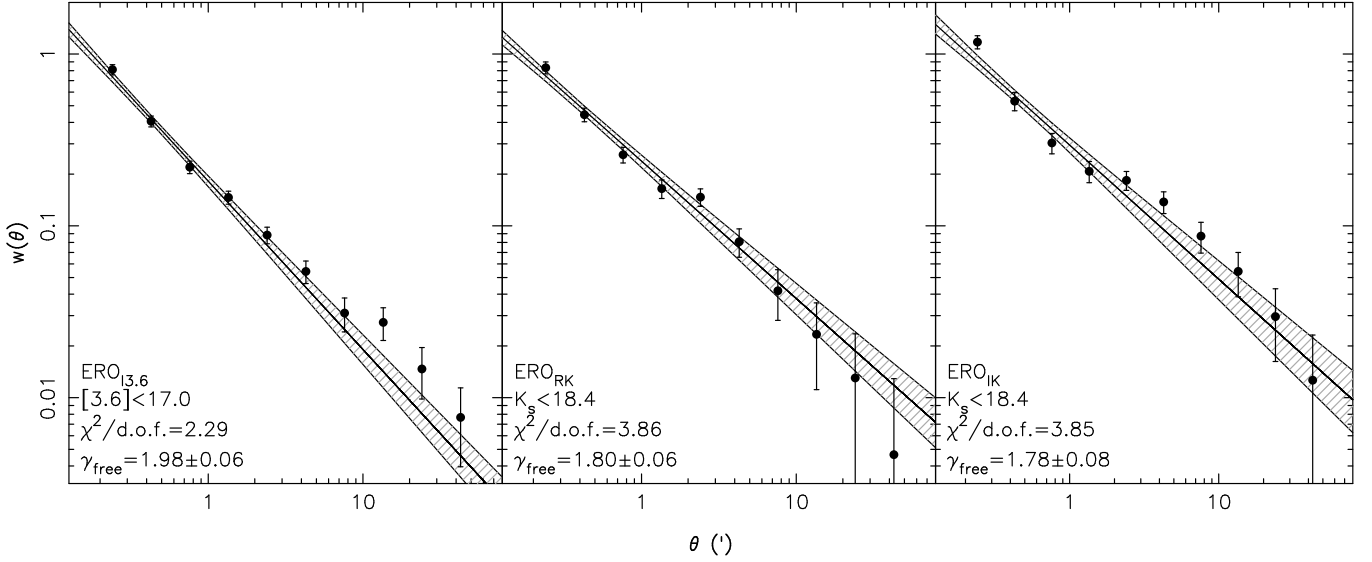}
\end{center}
\caption{Angular correlation functions $w(\theta)$. From left to right we show $w(\theta)$ for the ERO$_{I3.6}$ $[3.6]<17.0$, ERO$_{RK}$ $K_s<18.4$ and ERO$_{IK}$ $K_s<18.4$ samples. We also show the reduced $\chi^2$ and fitted power law $\gamma$ for each fit. These three samples are selected using different ERO criteria but show very similar clustering statistics. This suggests similar populations are being selected by all three ERO criteria.}
\label{EROacfs}
\end{figure*}

The angular correlation functions for the ERO$_{RK}$ $K_s<18.4$, ERO$_{IK}$ $K_s<18.4$ and ERO$_{I3.6}$ $[3.6]<17.0$ ERO samples are presented in Figure \ref{EROacfs}. Each of these ERO samples shows similar clustering statistics, which is an indication that similar galaxy populations are being included by each criteria. We then fit the angular correlation function for each sample using $\chi^2$ statistics including the full covariance matrix for the uncertainties and the functional form,
\begin{equation}
\label{acfpower}
w(\theta)=w(1')\left(\frac{\theta}{1'}\right)^{1-\gamma}.
\end{equation}
Many studies use a fixed value for $\gamma$ \citep[e.g.,][]{firth02,roche02,brown05,kong09,kim11} because they have insufficient data to constrain $\gamma.$ With our larger sample sizes ($N\gtrsim2000$), we are able to leave the power law $\gamma$ as a free parameter. To better compare with the prior literature and for our smaller samples with $1000\lesssim N<2000$, we also fit the correlation function with a fixed $\gamma=1.80$ \citep[e.g.,][]{kong09,kim11}. It should be noted that due to our field size, samples with $N<500$ do not produce reliable correlation functions and have been removed from our final results.

In many cases the reduced $\chi^2$ for our fits are distinctly greater than unity. We consider this an indication that a simple power law is not an adequate description of galaxy angular correlation functions. Some studies have tried to compensate for this with a double power law \citep[e.g.,][]{kim11}. However, this is again an empirical approximation and has no physical basis. For a robust analysis of galaxy correlation we must consider a more complex model of galaxy clustering such as galaxy-dark matter halo models \citep[e.g., the Halo Occupation Distribution;][]{selja00}, which we will address in a future paper.

%updated for 27/11/12 results

%updated for 27/11/12 results
\begin{figure}
\begin{center}
\resizebox{3.3in}{!}{\includegraphics{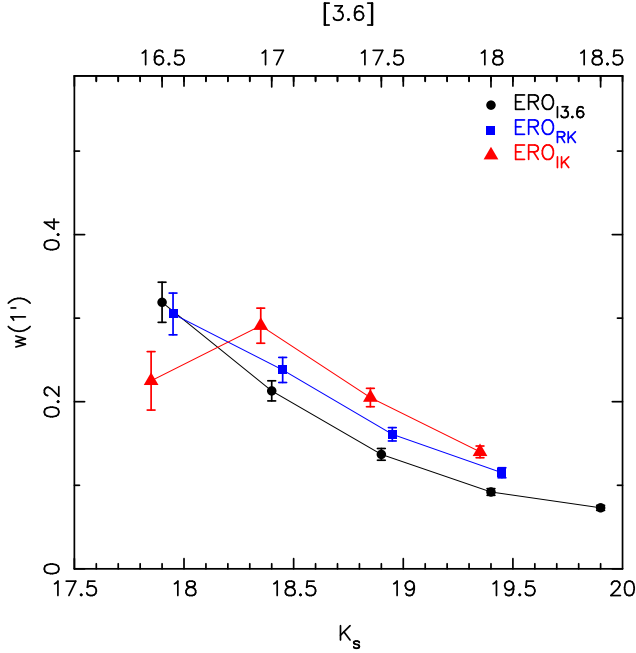}}
\end{center}
\caption{The angular correlation strength of our ERO samples as a function of apparent magnitude and for a fixed slope of $\gamma=1.80$. For clarity, magnitude offsets of $+0.05$ and $-0.05$ were applied to the ERO$_{I3.6}$ and ERO$_{IK}$ data, respectively. Angular clustering is seen to be strongly dependent on sample limiting magnitude. Here we have used $K_s\approx[3.6]+1.4$, based on the average of $K_s-[3.6]$ for our EROs, to place the $[3.6]$ samples on the figure.}
\label{WthetavL}
\end{figure}

%updated for 27/11/12 results
\begin{figure*}
\begin{center}
\includegraphics[scale=0.42]{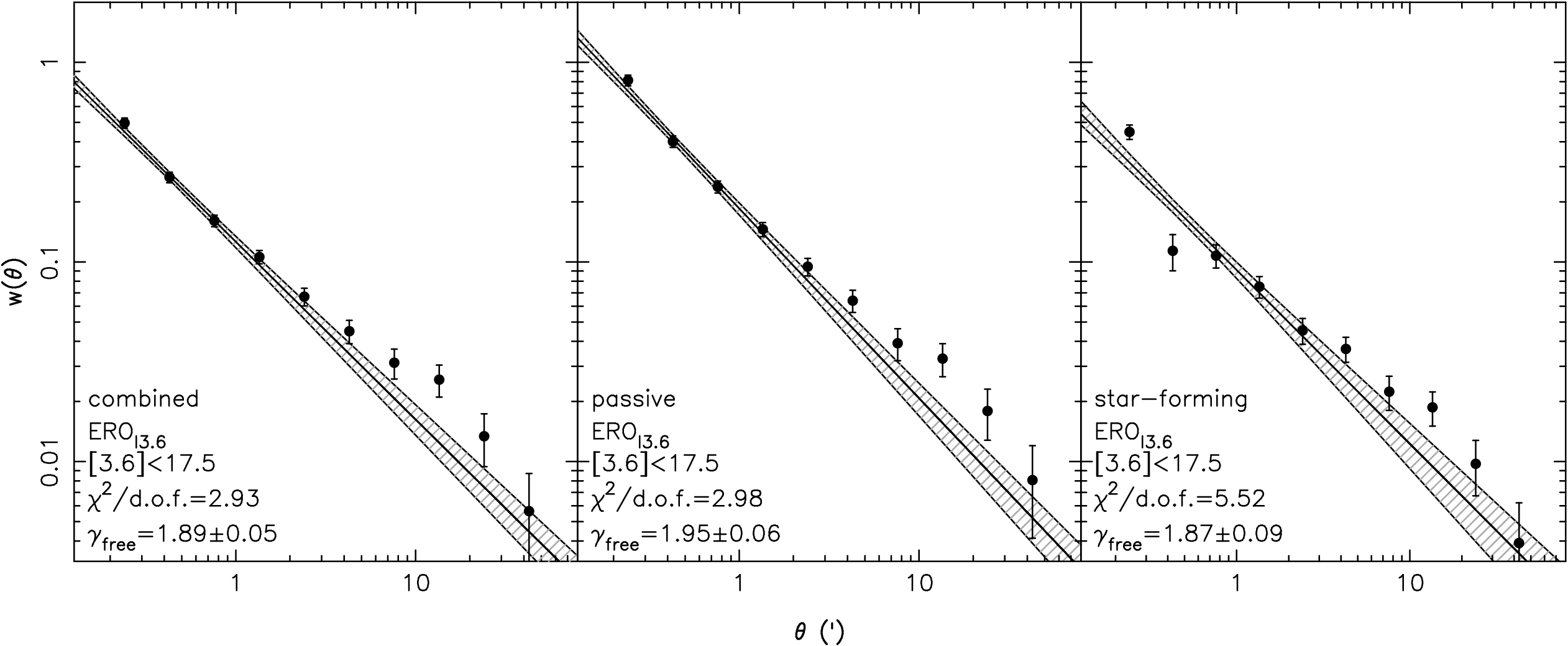}
\end{center}
\caption{Angular correlation functions $w(\theta)$ for (from left to right) the combined, passive and star-forming EROs brighter than $[3.6]<17.5$ selected with $(I-[3.6])>5.0$. We show the reduced $\chi^2$ and fitted power law $\gamma$ for each respective fit. Passive EROs show significantly stronger angular clustering strength than those with star-formation, especially on small angular scales.}
\label{EROacfs2}
\end{figure*}

The angular clustering of EROs is strongly dependent on limiting magnitude. As shown in Figure \ref{WthetavL}, the brightest sample for each ERO criterion has a $w(1\arcmin)$ clustering amplitude that is a factor of $\approx3$ stronger than the faintest sample (as before, we take $K_s\approx[3.6]+1.4$, based on the average of $K_s-[3.6]$ for our EROs). A complete summary of our combined ERO angular clustering analysis results is presented in Tables \ref{resultssum1} and \ref{resultssum3}. For ERO samples sub-divided by type, we observe stronger angular clustering for passive EROs in all three ERO samples. In all cases but the ERO$_{RK}$ $K_s<17.9$ sample, the $w(1\arcmin)$ clustering amplitude of the star-forming and passive ERO subsamples straddles that of the combined ERO sample. Figure \ref{EROacfs2} shows angular correlation functions for ERO$_{I3.6}$, $[3.6]<17.5$ combined and type-separated samples. There is no clear distinction between the passive, star-forming and mixed ERO$_{RK}$ and ERO$_{IK}$ samples, but the trend is the same as for the ERO$_{I3.6}$ sample. Separated sample results are presented in Tables \ref{resultssum2} and \ref{resultssum4}.

Freely fitted values for $\gamma$ were found to range between $\gamma=1.58-1.98$ for the combined ERO samples with a typical uncertainty of $\sigma_\gamma=0.05-0.10$. The passive EROs consistently yield a steeper power law index than the star-forming ERO populations (excepting the ERO$_{RK}$ $K_s<17.9$ sample), as can clearly be seen in Figure \ref{EROacfs2}. If we take the slope of $\gamma$ as an indicator of environment, where on the scale of $\approx 1$ Mpc a steep slope implies a more clustered distribution and a shallow $\gamma$ a smoother distribution, this implies that passive EROs typically occupy denser environments than star-forming EROs. 

%The redshift distribution - updated 30/10/12 
 \subsection{ERO redshift distribution}
 \label{dndz}
To transform the angular correlation function to a real-space correlation function we need to know the distribution of EROs in redshift. The redshift distribution (hereafter $dN/dz$) for each ERO sample was determined by number counts in bins of $\Delta z_{phot}=0.02$ over $0<z_{phot}<2$. The ANN$z$ code assigns galaxies with the same observed photometry to a particular $z_{phot}$, while the true $z_{spec}$ for these galaxies will be scattered above and below the $z_{phot}$ and in some cases catastrophically under/over-estimated (see Figure \ref{zzplot}). In our particular case, since the ERO cut selects a particular redshift distribution analogous to a redshift bin, we expect the true redshift distribution to be somewhat broader than given by the $dN/dz~$($z_{phot}$). This effect is clearly observed in Figure \ref{dndzcompare}. We therefore need to model the true $dN/dz~(z)$ based on the $dN/dz~$($z_{phot}$). 

We build our $dN/dz~(z)$ models in a two stage process. Firstly, we replace the $z_{phot}$ of each galaxy with a Gaussian probability distribution accounting for 90\% of each galaxy's distribution in redshift, where $\mu=z_{phot}$ and $\sigma$ is a function of $I-$band magnitude, $(B_W - R)$ color and $z_{phot}$. This accounts for the accuracy of the photometric redshift itself. Secondly, we account for catastrophic $z_{phot}$ errors with a second Gaussian probability distribution making up the remaining 10\% of each galaxy's distribution in redshift, where $\mu=z_{phot}$ and $\sigma=0.5$. This corresponds to the $\approx6\%$ of galaxies having $\Delta_z > 3\sigma_z z/(1+z)$, as observed in the data. The outliers need not be precisely modelled, as the \citet{limbe54} equation (see \S\ref{spatial}) depends on $(dN/dz)^2$. The choice of 10\% was made to better match the $z_{spec}$ distribution. Our resultant $dN/dz~(z)$ models are reasonable representations of the true redshift distribution, as shown by the examples in Figure \ref{dndzcompare}.

The $\bar{z}$ for each of our ERO samples is given in Tables \ref{resultssum1}, \ref{resultssum2}, \ref{resultssum3} and \ref{resultssum4}, and for the reasons discussed in \S\ref{photoz} we do not quote uncertainties on these values. As demonstrated by Figure \ref{dndzcompare}, the ERO$_{RK}$ sample has lower $\bar{z}$ than both the ERO$_{IK}$ and ERO$_{I3.6}$ samples, while the ERO$_{I3.6}$ and  ERO$_{IK}$ have similar $\bar{z}$. Fainter magnitude limited samples have higher $\bar{z}$, which is expected as a fainter limiting magnitude will include a larger number of more distant galaxies. The passive EROs and star-forming EROs in the type-separated samples have very similar $\bar{z}$ to each other. 

Reliable $z_{phot}$ are only assigned to objects with $I<23.5$, therefore not all of the EROs in our sample are assigned a $z_{phot}$. We account for this in two ways. For EROs with no $z_{phot}$, we assume that they have a $dN/dz~(z)$ similar to those with $z_{phot}$, and simply scale up $dN/dz~(z)$ to the full sample size. We do this for completeness, and it has little impact on the clustering results because the \citet{limbe54} equation (discussed in \S\ref{spatial}) only depends on the shape of $dN/dz~(z)$. We also provide in Tables \ref{resultssum3} and \ref{resultssum4} a complete clustering analysis including only objects with $z_{phot}$ information for comparison. In all but our faintest magnitude limited samples, $>50\%$ of EROs in each sample have $z_{phot}$ information and the modelled $dN/dz$ is considered a valid representation of the entire sample. 

It should be noted that when using smaller sample sizes but with an identical $dN/dz~(z)$, the clustering length of the $z_{phot}$ only samples (shown in Tables \ref{resultssum3} and \ref{resultssum4}), is observed to be systematically larger than that of samples including EROs without $z_{phot}$ information (shown in Tables \ref{resultssum1} and \ref{resultssum2}). For the majority of samples, the increase in $r_0$ is $\lesssim20\%$, with the exception of samples that have a large number of objects without $z_{phot}$ information.

We speculate that this is because objects without $z_{phot}$ information are likely faint and therefore more likely to have a higher average redshift. When included in the angular clustering measurement, these high-$z$ objects increase the number of uncorrelated pairs and will reduce the angular clustering amplitude. This would be accounted for if the redshift distribution was broader to account for the higher redshift of these faint objects. We note that samples with and without complete $z_{phot}$ information show the same general trends. Since we get better clustering statistics with greater sample size, we use our samples including EROs without $z_{phot}$ for discussion and the samples using only objects with a measured $z_{phot}$ are included as a cross-check.
 
 %Spatial clustering - updated 27/11/12 
 \subsection{Spatial clustering}
 \label{spatial} 
The spatial correlation function $\xi(\textbf{r})$ is related to $w(\theta)$ by the \citet{limbe54} equation,
\begin{eqnarray}
\label{limber}
\lefteqn{\hspace{-7mm}w(\theta)= } \nonumber
& &\frac{}{}\int_0^{\infty}{\frac{dN}{dz}\left[\int_0^{\infty}{\xi(\textbf{r}(\theta,z,z'),z)\frac{dN}{dz'}dz'}\right]dz}\\
& &\hspace{20mm}\left.\middle/\frac{}{}\left(\int_0^{\infty}{\frac{dN}{dz}dz}\right)^2\right..\\
\nonumber
\end{eqnarray}
The characteristic clustering length $r_0$ gives a measure of the clustering strength, and is related to $\xi(\textbf{r})$ by its functional form
\begin{equation}
\label{SCF}
\xi(\textbf{r})=\left(\frac{\textbf{r}}{r_0}\right)^{-\gamma}
\end{equation}
where $r_0$ is a measure of the spatial separation at which we expect to find an excess pair of galaxies above a random galaxy distribution, and $\gamma$ is as previously measured from the angular correlation function. The results of our spatial clustering analysis are given in Tables \ref{resultssum1} and \ref{resultssum3} for the combined ERO samples and in Tables \ref{resultssum2} and \ref{resultssum4} for the separated ERO samples. The uncertainty in $r_0$ was determined from the spread in $r_0$ values of the angular correlation function models with $\chi^2\leqslant(\chi^2_{min}+1)$. 

\begin{figure} 
\begin{center}
\resizebox{3.3in}{!}{\includegraphics{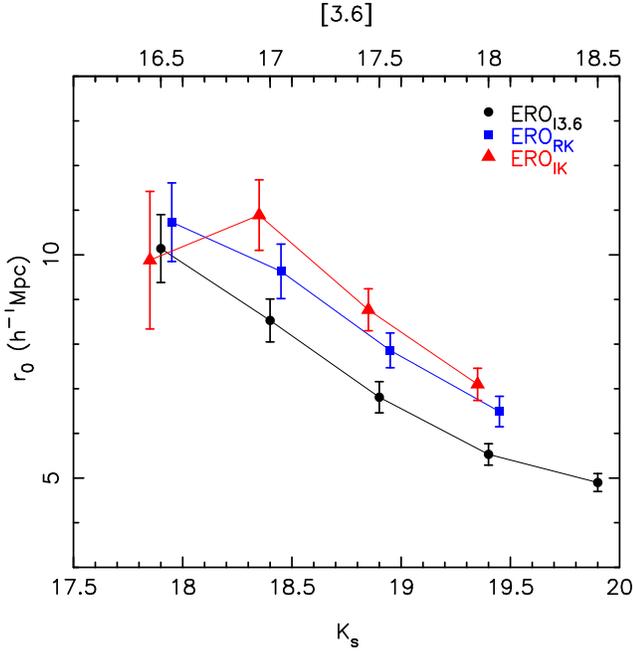}}
\end{center}
\caption{The real space spatial clustering strength of EROs as a function of apparent magnitude with fixed $\gamma=1.80$ for the three ERO samples. For clarity, magnitude offsets of $+0.05$ and $-0.05$ have been applied to the ERO$_{I3.6}$ and ERO$_{IK}$ data, respectively. The spatial clustering is strongly dependent on sample limiting magnitude. Here we assume $K_s\approx[3.6]+1.4$, based on the average of $K_s-[3.6]$ color of our EROs.}
\label{r0vL}
\end{figure}

Using our magnitude-limited samples we explore the spatial clustering of EROs as a function of limiting luminosity. For clarity and ease of comparison to previous studies we will only discuss results using fixed $\gamma=1.80$. In Figure \ref{r0vL} it is clear that ERO clustering strength is a strong function of apparent magnitude, since EROs occupy a particular redshift range, magnitude is a rough proxy for stellar mass. The ERO$_{RK}$ and ERO$_{IK}$ samples show stronger clustering than the ERO$_{I3.6}$ samples at similar magnitude limits (as before, we take $K_s\approx[3.6]+1.4$, based on the average of $K_s-[3.6]$ for our EROs). This could be due to the ERO$_{I3.6}$ sample having a larger star-forming galaxy fraction. 

For the ERO$_{RK}$ and ERO$_{IK}$ samples at a fixed magnitude limit, the clustering strength depends on the $\bar{z}$ of each sample, in agreement with previous work \citep[e.g.,][]{kim11}. This difference is easily accounted for because the difference in $\bar{z}$ between these two samples makes the objects in the ERO$_{IK}$ sample on average $\approx6\%$ brighter for a given limiting magnitude than the ERO$_{RK}$ sample. With this in mind we speculate that there is very little evolution in $r_0$ for roughly equivalent mass EROs between $z=1.2$ and $z=1.0$. The difference between the ERO$_{I3.6}$ sample and the ERO$_{RK}$ and ERO$_{IK}$ samples, we attribute (as above) to the difference in relative star-forming to passive ERO fractions. 

The brightest ERO$_{I3.6}$, ERO$_{RK}$ and ERO$_{IK}$ EROs have clustering lengths of $r_0=10.14\pm0.76~h^{-1}~\rm{Mpc}$, $r_0=10.73\pm0.88~h^{-1}~\rm{Mpc}$ and $r_0=9.88\pm1.54~h^{-1}~\rm{Mpc}$, respectively, comparable to that of $\approx4L^*$ local ellipticals \citep[SDSS; e.g.,][]{zehav10}. The faintest of ERO$_{RK}$ and $ERO_{IK}$ samples are still strongly clustered, with $r_0=6.49\pm0.34~h^{-1}~\rm{Mpc}$ and $r_0=7.10\pm0.36~h^{-1}~\rm{Mpc}$, respectively. This is comparable to that of  $2L^*$ local ellipticals. With a significantly higher star-forming fraction, the faintest ERO$_{I3.6}$ sample shows $r_0=4.90\pm0.20~h^{-1}~\rm{Mpc}$, similar to $\approx L^*$  local galaxies \citep[SDSS; e.g.,][]{zehav10}. It is clear that a larger star-forming fraction leads to reduced clustering strength. Since the star-forming fraction increases with apparent magnitude, this likely contributes to the observed decline of the combined ERO sample clustering length with apparent magnitude seen in Figure \ref{r0vL}.

The strong clustering of the faint ERO$_{RK}$ and ERO$_{IK}$ samples with respect to the faint ERO$_{I3.6}$ samples suggests that there is a dependence of clustering strength on ERO type. The clustering results for our separated ERO$_{RK}$ and ERO$_{IK}$ $K_s<17.9$ samples are largely inconclusive due to statistical uncertainties. However, all the type-separated ERO samples show a difference between the clustering of passive and star-forming EROs. In the ERO$_{I3.6}$ samples, which have the smallest uncertainties, there is a stark difference in the clustering between the two ERO types, with the clustering length $r_0$ of the passive EROs being a factor of $\approx1.4$ greater than that for the star-forming EROs at all 3 magnitude limits. It is clear that star-forming EROs occupy very different environments from passive EROs with comparable apparent magnitudes, and this is discussed further in \S\ref{link}. 

As a cross-check for the reliability of our ERO separation technique (\S\ref{separation}), we tested in the ERO$_{I3.6}$ sample out to $[3.6]<17.5$, the clustering of the $9,449$ $24~\mu$m non-detections. For a fixed $\gamma=1.80$ we estimate $r_0=8.46\pm0.48~h^{-1}\rm{Mpc}$, and this agrees well with the clustering length of the corresponding passive sample. We then tested the clustering of the $12,344$ $24~\mu$m detections for a fixed $\gamma=1.80$ and estimate $r_0=6.22\pm0.40~h^{-1}\rm{Mpc}$, which is similarly in agreement with the clustering length of the corresponding star-forming sample. Both results are slightly higher than for the color separated EROs. We take this as evidence that color classified passive EROs with $24~\mu$m detections increase the measured clustering length of the sample with $24~\mu$m detections, while star-forming EROs are absent in the sample with no $24~\mu$m detections and hence there is less likelihood of contamination. This conclusion is supported by the majority of color classified passive EROs with $24~\mu$m detections having strong D4000, as seen in Figures \ref{EROsepEGS_Ks} and \ref{EROsepEGS_CH1}. This confirms ERO type definitions based on $24~\mu$m colors as a robust method for separating star-forming and passive EROs.  

%updated for 27/11/12 results

%links to other redshifts - updated 27/11/12 
\subsection{Linking EROs throughout $z$ via spatial clustering}
\label{link}
The clustering of passive and star-forming EROs with comparable apparent magnitude show that the two ERO sub-populations occupy very different environments. This suggests that these sub-populations of EROs are the progenitors of different local Universe galaxies. In order to make links between EROs and galaxy populations at other epochs, we present comparisons of clustering for passive and star-forming $[3.6]<16.5$ EROs with samples of some of the largest galaxy clustering studies over $0<z<2.5$. 

\begin{figure}
\begin{center}
\resizebox{3.3in}{!}{\includegraphics{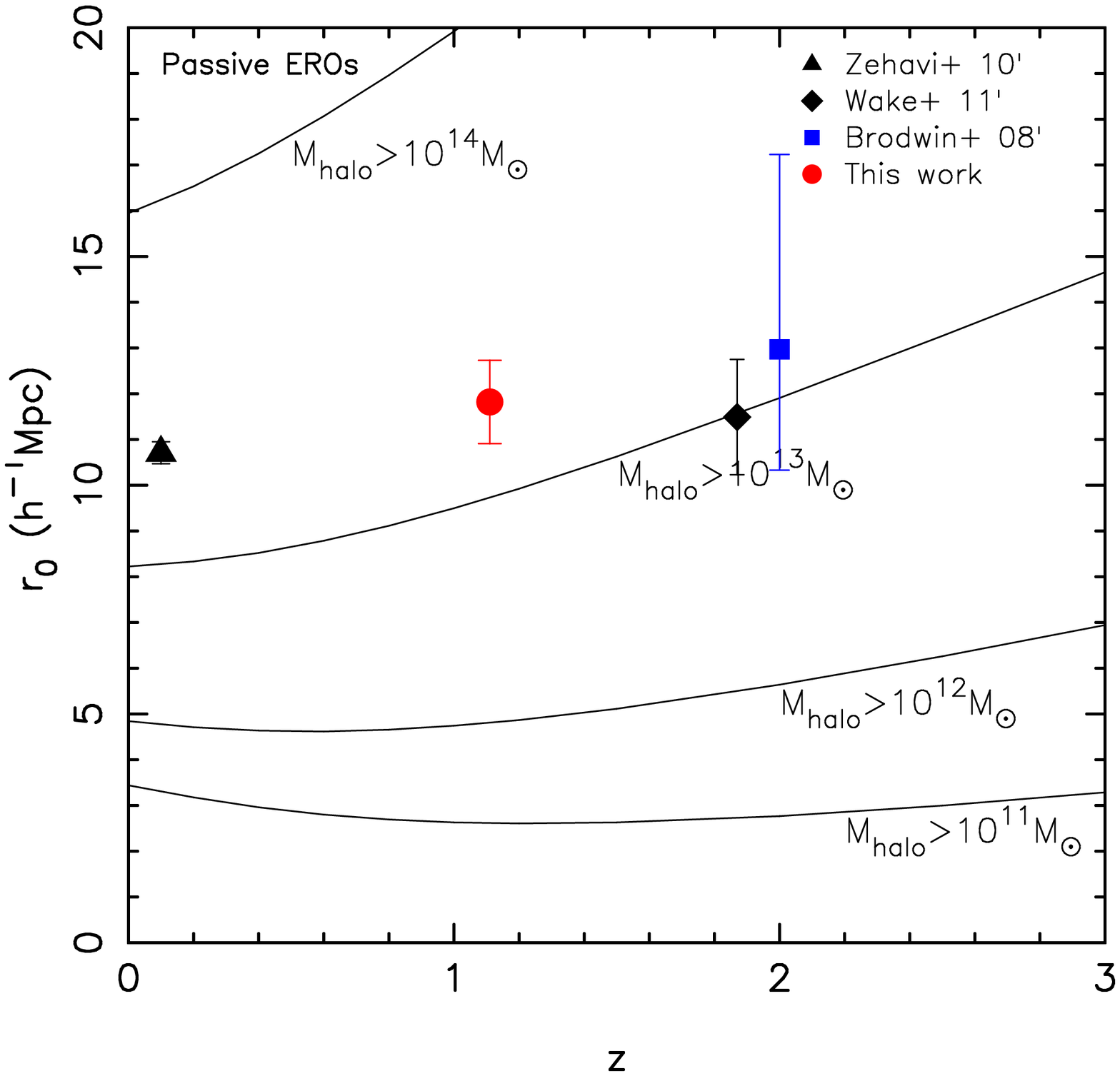}}
\end{center}
\caption{Spatial clustering as a function of redshift for the passive ERO$_{I3.6}$ $[3.6]<16.5$ sample (red circle; with fixed $\gamma$ for ease of comparison). The \citet{zehav10}, $M_{r}^{max}=-22.0$ sample is indicated by a triangle, the \citet{wake11}, stellar mass of $=10^{11}~\rm{M}_{\scriptsize{\odot}}$ sample with a diamond and the most luminous \citet{brodw08} dust obscured galaxy sample by the blue square. The curves show the evolution of $r_0(z)$ for fixed minimum halo mass, based on the methodology discussed in \S\ref{halomass}.}
\label{r0vz1}
\end{figure}

For ERO$_{I3.6}$ $[3.6]<16.5$ passive EROs we use the very brightest samples from each available comparison study and place our results in this context in Figure \ref{r0vz1}. We have also determined the expected $r_0(z)$ for dark matter halo samples and a range of minimum halo masses based on the methodology outlined in \S\ref{halomass}. The clustering at fixed minimum halo mass declines with redshift for massive halos (e.g.$~M_{halo}>10^{13} ~\rm{M}_{\scriptsize{\odot}})$. Halo masses however, grow with time and we would expect $r_0$ for a single, massive, growing halo to be slightly increasing with decreasing redshift. This is also the case for the massive galaxies at their cores. Figure \ref{r0vz1} shows that reasonably bright passive EROs are the likely progenitors of $\gtrsim4L^*$ local Universe ellipticals \citep{zehav10}. This suggests that bright EROs are the likely descendants of luminous dust-obscured galaxies (DOGs) at $z\approx2$ \citep{brodw08} and the likely progenitors of some of the most massive galaxies in the local Universe \citep{zehav10}. 

\begin{figure}
\begin{center}
\resizebox{3.3in}{!}{\includegraphics{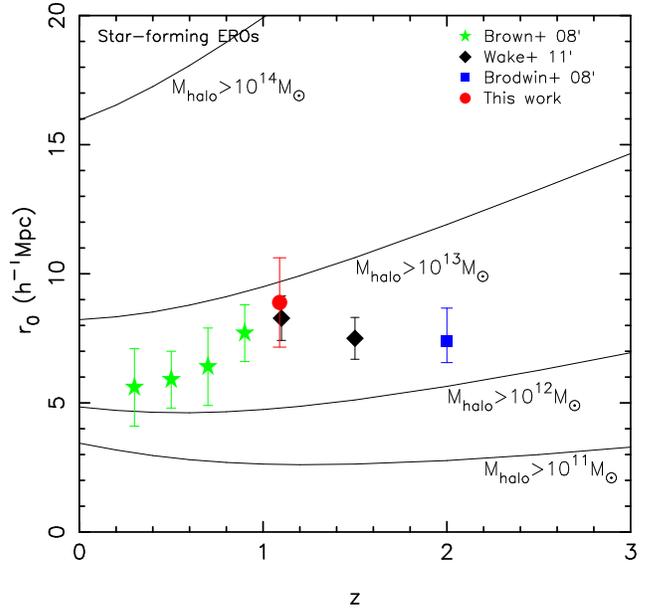}}
\end{center}
\caption{Spatial clustering as a function of redshift for the star-forming ERO$_{I3.6}$ [3.6]<17.5 sample (red circle; with fixed $\gamma$ for ease of comparison). Here we compare to the brightest absolute magnitude limited samples of passive red galaxies from \citet{brown08}, for each redshift range (green stars). The most massive, stellar mass samples from \citet{wake11}, for each redshift range (diamonds), and the least luminous dust obscured galaxy sample from \citet{brodw08} (blue square). The curves show the evolution of $r_0(z)$ for fixed minimum halo mass, based on the methodology discussed in \S\ref{halomass}.}
\label{r0vz2}
\end{figure}

In the case of our ERO$_{I3.6}$ $[3.6]<16.5$ star-forming EROs, their clustering strength of $r_0=8.89\pm1.73~h^{-1}\rm{Mpc}$, associates them with $\gtrsim2L^*$ local red galaxies \citep[e.g.,][]{zehav10}. Furthermore, when placed into context with comparison studies (Figure \ref{r0vz2}),  the expected evolution in clustering for the ERO$_{I3.6}$ $[3.6]<16.5$ star-forming EROs makes these likely progenitors of $z<0.8$ passive galaxies. Furthermore, the clustering length of $[3.6]<16.5$ star-forming EROs is comparable to that of passive $[3.6]<17.5$ EROs. This indicates the possibility of a duty cycle, whereby star-forming EROs turn into passive EROs with significantly fainter apparent magnitudes, which builds a consistent picture. This is evidence that $\gtrsim L^*$ local red galaxies are building up stellar mass at $z\approx1.2$. The $24~\mu$m flux limit of MAGES is consistent with that of Luminous Infra-Red galaxy (LIRG) emission at $z\approx1.2$ \citep{leflo05}, which sets the minimum star formation rate of at $\approx10 ~\rm{M}_{\scriptsize{\odot}}$ year$^{-1}$ \citep{noesk07} for star-forming EROs with $[3.6]<17.5$. These galaxies may be undergoing merger-induced star formation or are early forms of present-day ellipticals with continuing/residual star formation.

The above associations provide a strong indication that the ERO population is an earlier form of the elliptical population we observe in the local Universe. In general, the $r_0$ values for star-forming and passive EROs straddle that of the overall ERO population of the same apparent magnitude, with passive EROs being more strongly clustered than star-forming EROs for a particular magnitude limit. 

%- updated 28/09/12 
\subsection{Host halo mass} 
\label{halomass}
As galaxies are biased tracers of dark matter, it is possible to get an estimate of their host halo mass through a direct comparison of their clustering with that of dark matter halos. We determine the clustering of dark matter halos from the non-linear dark matter power spectrum. The power spectrum of dark matter is generated using the transfer function of \citet{eisen98}, the halo mass function and bias of \citet{sheth99}, and the halo model of \citet{selja00}. We assume NFW dark matter halo profiles \citep{navar96} with the concentration scalings of \citet{bullo01}. We adopt a simple halo occupation scheme assuming one galaxy per halo. We assume that the EROs represent the spatial distribution of their host halos on the scales of $\approx$Mpc, and determine their host halo mass by matching the $r_0$ of the EROs with that of mass limited dark matter halo samples.

This scheme provides crude estimates of the host halo mass, as it neglects the possibility of ERO satellite galaxies, and the correlation function of galaxies is not well described by a power law. A more accurate treatment may be achieved by halo occupation distribution modeling \citep[e.g.,][]{selja00,peaco00,zheng04,zehav04}, which we will address in a future paper. We show the corresponding halo mass estimates for combined ERO samples in Tables \ref{resultssum1} and \ref{resultssum3} and the separated ERO samples in Tables \ref{resultssum2} and \ref{resultssum4}. We find our brightest EROs are associated with dark matter halos with masses of the order $10^{13}M_{\scriptsize{\odot}}$. This host halo mass agrees with that predicted from the halo occupation distribution by \citet{gonza11}. The faintest EROs are associated with dark matter halos of $\approx10^{12}M_{\scriptsize{\odot}}$. These halo masses agree with the ERO-local galaxy population associations made using the spatial clustering measurements in the previous section.

%- updated 26/11/12 
\section{Discussion \& Summary}
\label{discussion}

We present the most statistically robust measurement of ERO clustering to date. The effects of cosmic variance on our results are modest considering that the integral constraint increases $w(\theta)$ by only $\approx3\%$ due to our large survey area. Our redshift distribution was derived from photometric redshifts, which are constrained with spectroscopy, convolved with their photometric uncertainties, and account for catastrophic $z_{phot}$ outliers, making them reliable representations of $dN/dz~(z)$. We find clear trends of decreasing clustering strength with apparent magnitude, which roughly translates to stellar mass since the redshifts are similar. Previous studies have hinted at such a relationship, but the samples were too small to draw solid conclusions \citep[e.g.,][]{brown05}. Our ERO$_{RK}$ results agree with our previous measurement of ERO clustering in a sub-field of the NDWFS \citep[][]{brown05}. We find that our measured ERO$_{RK}$ and ERO$_{IK}$ clustering lengths are typically lower than those of the previous largest area ERO studies by \citet{kim11} and \citet{kong09}, but agree within $2\sigma$. It should be noted that both \citet{kim11} and \citet{kong09} have likely underestimated their uncertainties, as they are based on Poisson uncertainties and assume a diagonal covariance matrix, so the true level of disagreement is smaller.

We tested whether the difference in ERO clustering results can be explained by the different criteria used to define ERO samples. We have clearly shown that the various ERO criteria select similar but not identical galaxy populations. Differences in number density, $\bar{z}$ and star-forming to passive ERO fraction are likely the main causes for differing estimates of $r_0$ between selection criteria. The results of ERO clustering studies using different selection criteria should be directly comparable so long as these factors are considered. For example, ERO$_{IK}$ $K_s<18.4$ EROs possess slightly stronger clustering properties to ERO$_{RK}$ $K_s<18.4$ EROs, because the difference in $\bar{z}$ between the samples means that the ERO$_{IK}$ EROs are systematically brighter than the ERO$_{RK}$ EROs. In the case of ERO$_{I3.6}$ we found that this criterion selected most of the same galaxies as the ERO$_{RK}$ and ERO$_{IK}$ samples, but also a greater surface density of objects for a comparable magnitude. The additional objects included a larger fraction of star-forming EROs than for the ERO$_{RK}$ and ERO$_{IK}$ samples, and this sample then showed systematically lower clustering for similar magnitude limits. This means that when you are trying to link EROs to local galaxy populations by their clustering, you cannot obtain meaningful results unless you separate the passive and star-forming EROs into separate samples. 

Our separation criteria, based on the work of \citet{stern06}, shows strong separation in both $(K_s-[24])$ and $([3.6]-[24])$ colors between EROs with star formation and those without. The success of this technique is its ability to pick out PAH emission from hot dust at high-redshift and over the redshift range required for ERO samples \citep{stern06,treye10}. The $([3.6]-[8.0])$ color is less effective because $8.0~\mu$m is no longer sampling strong PAH features at $z\gtrsim1$. Likewise the $(J-K)$ color is not useful for most EROs, because it does not bracket or straddle the $4000$\AA~break until $z\gtrsim1.4$. Deeper $24~\mu$m data would allow this technique to be used for fainter samples than analysed in this work. 

At fixed apparent magnitude, we find that the two ERO sub-populations occupy different environments, with passive EROs being more strongly clustered than those with star formation. The clustering strength for each type-separated ERO sample is also a function of apparent magnitude. In many cases, passive EROs have similar clustering properties to star-forming EROs of brighter apparent magnitude. This is consistent with a duty cycle, whereby star-forming EROs turn into passive EROs with significantly fainter apparent magnitudes. 

The SSFR seems to be lower in dense environments at low redshift \citep[e.g.][]{zehav10}, but some studies suggest that this trend changes at $z\gtrsim1.3$ \citep[e.g.,][]{coope07} to $z\approx2$ \citep[e.g.,][]{brodw08,magli08}. While we observe that EROs without star formation typically occupy denser environments than star-forming EROs, we see evidence for star formation in much denser environments than observed at $z<1$. This suggests that a transition between over-densities being dominated by passive galaxies to being dominated by star-forming galaxies could well be occurring at $z\gtrsim1.3$, in agreement with the findings of \citet{coope07} and \citet{brodw08}. 

Denser environments imply higher mass dark matter halos, and star-forming EROs seem to have host halo masses of up to $M_{halo}\approx10^{13}~\rm{M}_{\scriptsize{\odot}}$, an order of magnitude greater than the host halo mass of a typical star-forming galaxy at $z=0$ \citep{zheng07}. The shifting of the peak in star formation to higher mass halos at $z>1$ is in agreement with the results of previous high-redshift clustering studies \citep[e.g.,][]{magli08,wake11}. This could be evidence for "downsizing", where the host halo mass and stellar mass of star-forming galaxies decreases with decreasing redshift. Since a fixed halo mass at high redshift is more biased with respect to the underlying matter distribution \citep[e.g.,][]{sheth99,selja00} and therefore has a greater merger and mass accretion rate than at $z<1$, this result is not unexpected. 

The most significant result of our study is that the passive and star-forming ERO sub-populations are two distinct galaxy populations. Separating the ERO population into its sub-populations has allowed us to place EROs into an evolutionary sequence. Bright passive EROs are the likely progenitors of local $\gtrsim4L^*$ red galaxies, indicating that massive red galaxies have had their star formation truncated prior $z\approx1.2$. We were not able to obtain a clustering measurement for the progenitors of central cluster galaxies as such structures are rare, and as a consequence there are too few of these galaxies to measure their clustering. However, our brightest ERO samples certainly hint that the very brightest EROs are the likely progenitors of the most massive ellipticals in the local Universe. 

Interestingly, the clustering of star-forming EROs connects them as the likely progenitors of local $\gtrsim L^*$ red galaxies. It may seem peculiar that we detect significant star formation \citep[$\gtrsim10~\rm{M}_{\scriptsize{\odot}}/yr$; ][]{noesk07} at $z\approx1.2$ in progenitors of local $\gtrsim L^*$ ellipticals. As it is well established that star formation is expected to have ceased in these objects at $z\approx1.5$ \citep[e.g., gravitational lensing;][]{rusin05}. However, red/elliptical galaxies can be modelled with tau models with e-folding times of $\tau\lesssim1$Gyr and formation redshifts of $z\gtrsim2$. It is thus possible to have a galaxy where the bulk of star formation took place at $z\gtrsim1.5$, but at $z=1.2$ the SFR is still $\approx10~\rm{M}_{\scriptsize{\odot}}~\rm{year}^{-1}$ while by $z=0.5$ the SFR is negligible. Therefore, we find that local red galaxies, down to $L^*$, in some cases were still building stellar mass via star formation at this epoch.

\begin{acknowledgments} 
This work is based in part on observations made with the Spitzer Space Telescope, Spitzer/IRAC and Spitzer/MIPS, which is operated by the Jet Propulsion Laboratory, California Institute of Technology, under a contract with NASA. This work is based in part on observations made with the Kitt Peak National Observatory (KPNO). This research was supported by the National Optical Astronomy Observatory, which is operated by the Association of Universities for Research in Astronomy (AURA), Inc., under a cooperative agreement with the National Science Foundation. We thank our colleagues on the NDWFS, SDWFS and MAGES teams. We would like to thank Renbin Yan for providing the DEEP2 EGS spectroscopic data used in this work, Mark Dickinson for providing the FIDEL EGS catalogue used in this work, Xu Kong for providing the COSMOS ERO surface density data from \citet{kong09}, and Chris Conselice for providing the DEEP2 ERO surface density data from \citet{conse08}. D.P.P. acknowledges support from an Australian postgraduate award (APA) and a J.L. William postgraduate award. M.J.I.B. acknowledges support from an Australian Research Council (ARC) Future Fellowship. 

\end{acknowledgments}

\bibliographystyle{apj}

%\bibliography{Article_references.bib}

%- updated 27/11/12 
\begin{table*}[ht]
\label{resultssum1}
\centering
\caption{Summary of combined ERO clustering results}
\begin{tabular}{cccccccccc}
\hline\hline
ERO cut & Mag. limit & ERO type & $N$ & $\bar{z}$ & $\gamma$ & $w(1\arcmin)$ & $\chi^2/d.o.f$ & $r_0~(h^{-1}~\rm{Mpc})$ & $M_{halo}$ $(\times10^{12}~\rm{M}_{\scriptsize{\odot}})$\\
\hline
%I-[3.6]>5.0 EROs
$I-[3.6]>5.0$ & $[3.6]<16.5$ & ALL & $3209$ & 1.11 & $1.91\pm0.08$ & $0.295\pm0.028$ & $8.48$ & $8.77\pm1.65$  & $6.7^{+4.8}_{-2.9}$ \\
$I-[3.6]>5.0$ & $[3.6]<17.0$ & ALL & $9537$ & 1.14 & $1.98\pm0.06$ & $0.182\pm0.013$ & $2.29$ & $6.78\pm0.95$  & $2.9^{+1.5}_{-1.0}$ \\
$I-[3.6]>5.0$ & $[3.6]<17.5$ & ALL & $21800$ & 1.16 & $1.89\pm0.05$ & $0.126\pm0.008$ & $2.93$ & $6.10\pm0.77$  & $2.1^{+1.0}_{-0.7}$ \\
$I-[3.6]>5.0$ & $[3.6]<18.0$ & ALL & $40969$ & 1.17 & $1.87\pm0.05$ & $0.085\pm0.005$ & $9.02$ & $5.08\pm0.54$  & $1.1^{+0.4}_{-0.4}$ \\
$I-[3.6]>5.0$ & $[3.6]<18.5$ & ALL & $64370$ & 1.17 & $1.76\pm0.02$ & $0.071\pm0.002$ & $22.92$ & $4.93\pm0.21$  & $1.0^{+0.2}_{-0.0}$ \\
%R-K_S>5.0 EROs
$R-K_s>5.0$ & $K_s<17.9$ & ALL & $2715$ & 1.05 & $1.77\pm0.07$ & $0.312\pm0.031$ & $2.95$ & $11.21\pm2.08$  & $14.3^{+9.9}_{-6.1}$ \\
$R-K_s>5.0$ & $K_s<18.4$ & ALL & $6801$ & 1.09 & $1.80\pm0.06$ & $0.238\pm0.020$ & $3.86$ & $9.63\pm1.57$  & $8.9^{+5.3}_{-3.5}$ \\
$R-K_s>5.0$ & $K_s<18.9$ & ALL & $15013$ & 1.12 & $1.75\pm0.06$ & $0.169\pm0.014$ & $3.46$ & $8.43\pm1.39$  & $5.9^{+3.6}_{-2.2}$ \\
$R-K_s>5.0$ & $K_s<19.4$ & ALL & $28724$ & 1.13 & $1.83\pm0.06$ & $0.112\pm0.008$ & $7.34$ & $6.26\pm0.93$  & $2.3^{+1.3}_{-0.9}$ \\
%I-K_s>4.0 EROs
$I-K_s>4.0$ & $K_s<17.9$ & ALL & $1314$ & 1.08 & $1.58\pm0.10$ & $0.235\pm0.041$ & $2.80$ & $12.68\pm3.14$  & $21.0^{+20.0}_{-12.3}$ \\
$I-K_s>4.0$ & $K_s<18.4$ & ALL & $3672$ & 1.15 & $1.78\pm0.08$ & $0.295\pm0.029$ & $3.85$ & $11.19\pm2.27$  & $14.2^{+11.0}_{-6.6}$ \\
$I-K_s>4.0$ & $K_s<18.9$ & ALL & $9829$ & 1.18 & $1.71\pm0.07$ & $0.223\pm0.022$ & $6.92$ & $9.98\pm2.00$  & $10.0^{+7.6}_{-4.4}$ \\
$I-K_s>4.0$ & $K_s<19.4$ & ALL & $22451$ & 1.19 & $1.75\pm0.07$ & $0.148\pm0.014$ & $9.83$ & $7.62\pm1.35$  & $4.2^{+2.6}_{-2.0}$ \\
\hline
%I-[3.6]>5.0 EROs fixed gamma
$I-[3.6]>5.0$ & $[3.6]<16.5$ & ALL & $3209$ & 1.11 & $1.80$ & $0.319\pm0.024$ & $8.21$ & $10.14\pm0.76$  & $10.5^{+2.6}_{-2.2}$ \\
$I-[3.6]>5.0$ & $[3.6]<17.0$ & ALL & $9537$ & 1.14 & $1.80$ & $0.213\pm0.012$ & $3.09$ & $8.53\pm0.48$  & $6.1^{+1.1}_{-1.0}$ \\
$I-[3.6]>5.0$ & $[3.6]<17.5$ & ALL & $21800$ & 1.16 & $1.80$ & $0.137\pm0.007$ & $2.73$ & $6.81\pm0.35$  & $3.0^{+0.5}_{-0.4}$ \\
$I-[3.6]>5.0$ & $[3.6]<18.0$ & ALL & $40969$ & 1.17 & $1.80$ & $0.092\pm0.004$ & $7.69$ & $5.53\pm0.24$  & $1.5^{+0.2}_{-0.2}$ \\
$I-[3.6]>5.0$ & $[3.6]<18.5$ & ALL & $64370$ & 1.17 & $1.80$ & $0.073\pm0.003$ & $13.90$ & $4.90\pm0.20$  & $1.0^{+0.1}_{-0.1}$ \\
%R-K_s>5.0 EROs fixed gamma
$R-K_s>5.0$ & $K_s<17.9$ & ALL & $2715$ & 1.05 & $1.80$ & $0.305\pm0.025$ & $3.02$ & $10.73\pm0.88$  & $12.5^{+3.3}_{-2.9}$ \\
$R-K_s>5.0$ & $K_s<18.4$ & ALL & $6801$ & 1.09 & $1.80$ & $0.238\pm0.015$ & $3.86$ & $9.63\pm0.61$  & $8.9^{+1.7}_{-1.6}$ \\
$R-K_s>5.0$ & $K_s<18.9$ & ALL & $15013$ & 1.12 & $1.80$ & $0.161\pm0.008$ & $3.94$ & $7.86\pm0.39$  & $4.7^{+0.8}_{-0.7}$ \\
$R-K_s>5.0$ & $K_s<19.4$ & ALL & $28724$ & 1.13 & $1.80$ & $0.115\pm0.006$ & $7.03$ & $6.49\pm0.34$  & $2.9^{+1.5}_{-1.0}$ \\
%I-K_s>4.0 EROs fixed gamma
$I-K_s>4.0$ & $K_s<17.9$ & ALL & $1314$ & 1.08 & $1.80$ & $0.225\pm0.035$ & $3.54$ & $9.88\pm1.54$  & $9.7^{+5.5}_{-3.9}$ \\
$I-K_s>4.0$ & $K_s<18.4$ & ALL & $3672$ & 1.15 & $1.80$ & $0.291\pm0.021$ & $3.94$ & $10.89\pm0.79$  & $13.1^{+3.1}_{-2.7}$ \\
$I-K_s>4.0$ & $K_s<18.9$ & ALL & $9829$ & 1.18 & $1.80$ & $0.205\pm0.011$ & $8.72$ & $8.77\pm0.47$  & $6.7^{+1.2}_{-1.1}$ \\
$I-K_s>4.0$ & $K_s<19.4$ & ALL & $22451$ & 1.19 & $1.80$ & $0.140\pm0.007$ & $11.85$ & $7.10\pm0.36$  & $3.4^{+0.6}_{-0.4}$ \\
\hline\hline
\end{tabular}
\\
Note: Clustering results for samples with $N<500$ are not shown due to poor quality statistics for our field area.
\end{table*}

\begin{table*}[ht]
\label{resultssum2}
\centering
\caption{Summary of type-separated ERO clustering results.}
\begin{tabular}{ccccccccccc}
\hline\hline
ERO &{Mag.}&{ERO}&{$N$}&{$\%$}&{$\bar{z}$}&{$\gamma$}&{$w(1\arcmin)$}&{$\chi^2/d.o.f$}&{$r_0$}&{$M_{halo}$}\\
{cut}&{limit}&{type}&{}&{total}&{}&{}&{}&{}&{$(h^{-1}~\rm{Mpc})$}&{$(\times10^{12}~\rm{M}_{\scriptsize{\odot}})$}\\
\hline
%I-[3.6]>5.0 EROs
$I-[3.6]>5.0$ & $[3.6]<16.5$ & E & $2281$ & 71 & 1.11 & $1.92\pm0.09$ & $0.419\pm0.041$ & $5.91$ & $10.01\pm1.79$  & $10.1^{+6.6}_{-4.7}$ \\
$I-[3.6]>5.0$ & $[3.6]<16.5$ & SF & $928$ & 29 & 1.09 & $1.82\pm0.14$ & $0.232\pm0.045$ & $7.83$ & $8.75\pm2.27$  & $6.6^{+7.0}_{-4.1}$ \\
$I-[3.6]>5.0$ & $[3.6]<17.0$ & E & $6183$ & 65 & 1.14 & $2.00\pm0.07$ & $0.265\pm0.020$ & $4.17$ & $7.73\pm1.05$  & $4.5^{+2.1}_{-1.7}$ \\
$I-[3.6]>5.0$ & $[3.6]<17.0$ & SF & $3354$ & 35 & 1.14 & $1.83\pm0.10$ & $0.139\pm0.016$ & $6.53$ & $7.05\pm1.45$  & $3.3^{+2.7}_{-1.7}$ \\
$I-[3.6]>5.0$ & $[3.6]<17.5$ & E & $12905$ & 59 & 1.16 & $1.95\pm0.06$ & $0.185\pm0.013$ & $2.98$ & $6.84\pm0.88$  & $3.0^{+1.4}_{-1.1}$ \\
$I-[3.6]>5.0$ & $[3.6]<17.5$ & SF & $8895$ & 41 & 1.16 & $1.87\pm0.09$ & $0.091\pm0.008$ & $5.52$ & $5.52\pm1.02$  & $1.5^{+1.1}_{-0.7}$ \\
%R-K_s>5.0 EROs
$R-K_s>5.0$ & $K_s<17.9$ & E & $2213$ & 82 & 1.04 & $1.89\pm0.08$ & $0.335\pm0.033$ & $3.02$ & $10.36\pm1.87$  & $11.2^{+7.5}_{-4.9}$ \\
$R-K_s>5.0$ & $K_s<17.9$ & SF & $502$ & 18 & 1.06 & $2.04\pm0.17$ & $0.401\pm0.075$ & $4.50$ & $9.20\pm2.70$  & $7.7^{+8.6}_{-5.2}$ \\
$R-K_s>5.0$ & $K_s<18.4$ & E & $4744$ & 70 & 1.08 & $1.85\pm0.06$ & $0.260\pm0.022$ & $5.56$ & $9.52\pm1.50$  & $8.6^{+5.0}_{-3.5}$ \\
$R-K_s>5.0$ & $K_s<18.4$ & SF & $2057$ & 30 & 1.11 & $1.79\pm0.09$ & $0.262\pm0.031$ & $1.97$ & $10.34\pm2.23$  & $11.1^{+9.3}_{-5.5}$ \\
%I-K_s>4.0 EROs
$I-K_s>4.0$ & $K_s<17.9$ & E & $1114$ & 85 & 1.07 & $1.73\pm0.12$ & $0.247\pm0.042$ & $2.37$ & $11.37\pm2.90$  & $14.9^{+15.2}_{-8.8}$ \\
$I-K_s>4.0$ & $K_s<17.9$ & SF & $200$ & 15 & 1.15 & $-$ & $-$ & $-$ & $-$  & $-$ \\
$I-K_s>4.0$ & $K_s<18.4$ & E & $2481$ & 68 & 1.13 & $1.78\pm0.09$ & $0.313\pm0.033$ & $5.14$ & $11.60\pm2.46$  & $15.9^{+12.9}_{-7.8}$ \\
$I-K_s>4.0$ & $K_s<18.4$ & SF & $1191$ & 32 & 1.18 & $1.61\pm0.10$ & $0.332\pm0.051$ & $3.27$ & $13.95\pm3.33$  & $28.1^{+26.2}_{-15.7}$ \\
\hline
%I-[3.6]>5.0 EROs fixed gamma
$I-[3.6]>5.0$ & $[3.6]<16.5$ & E & $2281$ & 71 & 1.11 & $1.80$ & $0.456\pm0.035$ & $6.00$ & $11.82\pm0.91$  & $16.9^{+4.3}_{-3.7}$ \\
$I-[3.6]>5.0$ & $[3.6]<16.5$ & SF & $928$ & 29 & 1.09 & $1.80$ & $0.231\pm0.045$ & $7.76$ & $8.89\pm1.73$  & $6.9^{+5.2}_{-3.4}$ \\
$I-[3.6]>5.0$ & $[3.6]<17.0$ & E & $6183$ & 65 & 1.14 & $1.80$ & $0.315\pm0.019$ & $4.86$ & $10.12\pm0.61$  & $10.4^{+2.0}_{-1.8}$ \\
$I-[3.6]>5.0$ & $[3.6]<17.0$ & SF & $3354$ & 35 & 1.14 & $1.80$ & $0.141\pm0.016$ & $6.38$ & $7.27\pm0.83$  & $3.7^{+1.4}_{-1.2}$ \\
$I-[3.6]>5.0$ & $[3.6]<17.5$ & E & $12905$ & 59 & 1.16 & $1.80$ & $0.212\pm0.011$ & $3.43$ & $8.29\pm0.43$  & $5.6^{+1.0}_{-0.9}$ \\
$I-[3.6]>5.0$ & $[3.6]<17.5$ & SF & $8895$ & 41 & 1.16 & $1.80$ & $0.095\pm0.007$ & $5.39$ & $5.91\pm0.44$  & $1.9^{+0.5}_{-0.4}$ \\
%R-K_s>5.0 EROs fixed gamma
$R-K_s>5.0$ & $K_s<17.9$ & E & $2213$ & 82 & 1.04 & $1.80$ & $0.357\pm0.030$ & $3.01$ & $11.77\pm0.99$  & $16.6^{+4.7}_{-4.0}$ \\
$R-K_s>5.0$ & $K_s<17.9$ & SF & $502$ & 18 & 1.06 & $1.80$ & $0.400\pm0.080$ & $4.42$ & $11.69\pm2.34$  & $16.3^{+12.1}_{-8.2}$ \\
$R-K_s>5.0$ & $K_s<18.4$ & E & $4744$ & 70 & 1.08 & $1.80$ & $0.271\pm0.019$ & $5.41$ & $10.23\pm0.72$  & $10.8^{+2.5}_{-2.1}$ \\
$R-K_s>5.0$ & $K_s<18.4$ & SF & $2057$ & 30 & 1.11 & $1.80$ & $0.261\pm0.027$ & $2.00$ & $10.21\pm1.06$  & $10.7^{+3.8}_{-3.1}$ \\
%I-K_s>4.0 EROs
$I-K_s>4.0$ & $K_s<17.9$ & E & $1114$ & 85 & 1.07 & $1.80$ & $0.246\pm0.040$ & $2.52$ & $10.56\pm1.72$  & $11.9^{+7.1}_{-5.1}$ \\
$I-K_s>4.0$ & $K_s<17.9$ & SF & $200$ & 15 & 1.15 & $-$ & $-$ & $-$ & $-$  & $-$ \\
$I-K_s>4.0$ & $K_s<18.4$ & E & $2481$ & 68 & 1.13 & $1.80$ & $0.309\pm0.027$ & $5.20$ & $11.29\pm0.99$  & $14.6^{+4.1}_{-3.6}$ \\
$I-K_s>4.0$ & $K_s<18.4$ & SF & $1191$ & 32 & 1.18 & $1.80$ & $0.321\pm0.042$ & $4.64$ & $11.17\pm1.46$  & $14.2^{+6.6}_{-4.9}$ \\
\hline\hline
\end{tabular}
\\ 
Note: Clustering results for samples with $N<500$ are not shown due to poor quality statistics for our field area.\\ 
*The ERO types are labeled as E for passive EROs and SF for star-forming EROs.\\ 
\% total is the fraction of combined EROs in a particular ERO-type sub-population.
\end{table*}

\begin{table*}[ht]
\label{resultssum3}
\centering
\caption{Summary of combined ERO clustering results (ONLY galaxies with $z_{phot}$ information).}
\begin{tabular}{cccccccccc}
\hline\hline
ERO cut & Mag. limit & ERO type & $N$ & $\bar{z}$ & $\gamma$ & $w(1\arcmin)$ & $\chi^2/d.o.f$ & $r_0~(h^{-1}~\rm{Mpc})$ & $M_{halo}$ $(\times10^{12}~\rm{M}_{\scriptsize{\odot}})$\\
\hline
%I-[3.6]>5.0 EROs
$I-[3.6]>5.0$ & $[3.6]<16.5$ & ALL & $2539$ & 1.11 & $1.89\pm0.07$ & $0.357\pm0.033$ & $3.25$ & $9.88\pm1.68$  & $9.7^{+6.1}_{-3.8}$ \\
$I-[3.6]>5.0$ & $[3.6]<17.0$ & ALL & $7464$ & 1.14 & $1.88\pm0.05$ & $0.254\pm0.019$ & $1.88$ & $8.74\pm1.21$  & $6.6^{+3.3}_{-2.3}$ \\
$I-[3.6]>5.0$ & $[3.6]<17.5$ & ALL & $15380$ & 1.16 & $1.76\pm0.06$ & $0.198\pm0.016$ & $1.56$ & $8.64\pm1.22$  & $6.3^{+2.9}_{-2.4}$ \\
$I-[3.6]>5.0$ & $[3.6]<18.0$ & ALL & $22754$ & 1.17 & $1.73\pm0.05$ & $0.158\pm0.012$ & $2.24$ & $7.89\pm1.18$  & $4.8^{+2.6}_{-1.7}$ \\
$I-[3.6]>5.0$ & $[3.6]<18.5$ & ALL & $25703$ & 1.17 & $1.76\pm0.05$ & $0.145\pm0.011$ & $2.80$ & $7.39\pm1.04$  & $3.9^{+2.0}_{-1.4}$ \\
%R-K_S>5.0 EROs
$R-K_s>5.0$ & $K_s<17.9$ & ALL & $2155$ & 1.05 & $1.68\pm0.07$ & $0.402\pm0.043$ & $3.68$ & $14.36\pm2.65$  & $30.7^{+21.1}_{-13.8}$ \\
$R-K_s>5.0$ & $K_s<18.4$ & ALL & $5395$ & 1.09 & $1.76\pm0.06$ & $0.312\pm0.027$ & $1.94$ & $11.67\pm1.85$  & $16.2^{+9.3}_{-6.5}$ \\
$R-K_s>5.0$ & $K_s<18.9$ & ALL & $10795$ & 1.12 & $1.69\pm0.05$ & $0.239\pm0.021$ & $1.87$ & $10.90\pm1.85$  & $13.1^{+8.2}_{-5.1}$ \\
$R-K_s>5.0$ & $K_s<19.4$ & ALL & $15209$ & 1.13 & $1.70\pm0.06$ & $0.205\pm0.017$ & $1.37$ & $9.80\pm1.45$  & $9.4^{+5.0}_{-3.7}$ \\
%I-K_s>4.0 EROs
$I-K_s>4.0$ & $K_s<17.9$ & ALL & $940$ & 1.08 & $1.90\pm0.13$ & $0.290\pm0.046$ & $1.96$ & $10.29\pm2.48$  & $11.0^{+10.3}_{-6.4}$ \\
$I-K_s>4.0$ & $K_s<18.4$ & ALL & $2653$ & 1.15 & $1.80\pm0.07$ & $0.353\pm0.034$ & $2.47$ & $12.12\pm2.22$  & $18.2^{+12.4}_{-7.8}$ \\
$I-K_s>4.0$ & $K_s<18.9$ & ALL & $6133$ & 1.18 & $1.73\pm0.06$ & $0.260\pm0.023$ & $3.09$ & $10.71\pm1.83$  & $12.4^{+7.8}_{-5.2}$ \\
$I-K_s>4.0$ & $K_s<19.4$ & ALL & $8988$ & 1.19 & $1.75\pm0.06$ & $0.217\pm0.018$ & $1.49$ & $9.48\pm1.56$  & $8.5^{+5.2}_{-3.2}$ \\
\hline
%I-[3.6]>5.0 EROs fixed gamma
$I-[3.6]>5.0$ & $[3.6]<16.5$ & ALL & $2539$ & 1.11 & $1.80$ & $0.381\pm0.030$ & $3.27$ & $11.19\pm0.88$  & $14.2^{+3.6}_{-3.2}$ \\
$I-[3.6]>5.0$ & $[3.6]<17.0$ & ALL & $7464$ & 1.14 & $1.80$ & $0.273\pm0.016$ & $1.91$ & $9.79\pm0.57$  & $9.4^{+1.8}_{-1.6}$ \\
$I-[3.6]>5.0$ & $[3.6]<17.5$ & ALL & $15380$ & 1.16 & $1.80$ & $0.190\pm0.010$ & $1.75$ & $8.16\pm0.43$  & $5.3^{+0.8}_{-0.8}$ \\
$I-[3.6]>5.0$ & $[3.6]<18.0$ & ALL & $22754$ & 1.17 & $1.80$ & $0.147\pm0.007$ & $2.79$ & $7.17\pm0.34$  & $3.5^{+0.6}_{-0.5}$ \\
$I-[3.6]>5.0$ & $[3.6]<18.5$ & ALL & $25703$ & 1.17 & $1.80$ & $0.139\pm0.006$ & $3.07$ & $7.00\pm0.30$  & $3.2^{+0.5}_{-0.4}$ \\
%R-K_s>5.0 EROs fixed gamma
$R-K_s>5.0$ & $K_s<17.9$ & ALL & $2155$ & 1.05 & $1.80$ & $0.364\pm0.031$ & $4.32$ & $11.84\pm1.01$  & $16.9^{+4.9}_{-4.0}$ \\
$R-K_s>5.0$ & $K_s<18.4$ & ALL & $5395$ & 1.09 & $1.80$ & $0.301\pm0.019$ & $2.12$ & $10.97\pm0.69$  & $13.4^{+2.8}_{-2.5}$ \\
$R-K_s>5.0$ & $K_s<18.9$ & ALL & $10795$ & 1.12 & $1.80$ & $0.214\pm0.012$ & $2.86$ & $9.21\pm0.52$  & $7.7^{+1.3}_{-1.3}$ \\
$R-K_s>5.0$ & $K_s<19.4$ & ALL & $15209$ & 1.13 & $1.80$ & $0.185\pm0.010$ & $2.18$ & $8.45\pm0.46$  & $6.6^{+3.3}_{-2.3}$ \\
%I-K_s>4.0 EROs fixed gamma
$I-K_s>4.0$ & $K_s<17.9$ & ALL & $940$ & 1.08 & $1.80$ & $0.297\pm0.048$ & $2.00$ & $11.53\pm1.86$  & $15.6^{+9.2}_{-6.6}$ \\
$I-K_s>4.0$ & $K_s<18.4$ & ALL & $2653$ & 1.15 & $1.80$ & $0.353\pm0.028$ & $2.47$ & $12.12\pm0.96$  & $18.2^{+4.8}_{-4.0}$ \\
$I-K_s>4.0$ & $K_s<18.9$ & ALL & $6133$ & 1.18 & $1.80$ & $0.243\pm0.016$ & $3.43$ & $9.64\pm0.63$  & $9.0^{+2.0}_{-1.6}$ \\
$I-K_s>4.0$ & $K_s<19.4$ & ALL & $8988$ & 1.19 & $1.80$ & $0.207\pm0.012$ & $1.73$ & $8.83\pm0.51$  & $6.8^{+1.2}_{-1.2}$ \\
\hline\hline
\end{tabular}
\\ 
Note: Clustering results for samples with $N<500$ are not shown due to poor quality statistics for our field area.\\ 
These samples contain only objects that have $z_{phot}$ information.
\end{table*}

\begin{table*}[ht]
\label{resultssum4}
\centering
\caption{Summary of type-separated ERO clustering results (ONLY galaxies with $z_{phot}$ information).}
\begin{tabular}{ccccccccccc}
\hline\hline
ERO &{Mag.}&{ERO}&{$N$}&{$\%$}&{$\bar{z}$}&{$\gamma$}&{$w(1\arcmin)$}&{$\chi^2/d.o.f$}&{$r_0$}&{$M_{halo}$}\\
{cut}&{limit}&{type}&{}&{total}&{}&{}&{}&{}&{$(h^{-1}~\rm{Mpc})$}&{$(\times10^{12}~\rm{M}_{\scriptsize{\odot}})$}\\
\hline
%I-[3.6]>5.0 EROs
$I-[3.6]>5.0$ & $[3.6]<16.5$ & E & $1849$ & 73 & 1.11 & $1.98\pm0.08$ & $0.500\pm0.045$ & $2.95$ & $10.34\pm1.79$  & $11.1^{+7.1}_{-4.5}$ \\
$I-[3.6]>5.0$ & $[3.6]<16.5$ & SF & $690$ & 27 & 1.09 & $1.57\pm0.09$ & $0.258\pm0.063$ & $6.83$ & $11.88\pm3.20$  & $17.1^{+18.6}_{-10.5}$ \\
$I-[3.6]>5.0$ & $[3.6]<17.0$ & E & $5003$ & 67 & 1.14 & $2.01\pm0.06$ & $0.330\pm0.024$ & $2.55$ & $8.56\pm1.11$  & $6.1^{+2.7}_{-2.2}$ \\
$I-[3.6]>5.0$ & $[3.6]<17.0$ & SF & $2461$ & 33 & 1.14 & $1.63\pm0.09$ & $0.188\pm0.026$ & $5.68$ & $9.95\pm2.17$  & $9.9^{+8.3}_{-5.1}$ \\
$I-[3.6]>5.0$ & $[3.6]<17.5$ & E & $9697$ & 63 & 1.16 & $1.90\pm0.05$ & $0.249\pm0.018$ & $1.28$ & $8.30\pm1.14$  & $5.6^{+2.8}_{-1.8}$ \\
$I-[3.6]>5.0$ & $[3.6]<17.5$ & SF & $5683$ & 37 & 1.16 & $1.88\pm0.08$ & $0.142\pm0.014$ & $7.01$ & $6.95\pm1.33$  & $3.2^{+2.4}_{-1.5}$ \\
%R-K_s>5.0 EROs
$R-K_s>5.0$ & $K_s<17.9$ & E & $1761$ & 82 & 1.04 & $1.92\pm0.08$ & $0.401\pm0.039$ & $3.49$ & $11.04\pm1.94$  & $13.6^{+8.9}_{-5.9}$ \\
$R-K_s>5.0$ & $K_s<17.9$ & SF & $394$ & 18 & 1.06 & $-$ & $-$ & $-$ & $-$  & $-$ \\
$R-K_s>5.0$ & $K_s<18.4$ & E & $3804$ & 71 & 1.08 & $1.89\pm0.07$ & $0.327\pm0.028$ & $3.15$ & $10.35\pm1.56$  & $11.2^{+5.9}_{-4.5}$ \\
$R-K_s>5.0$ & $K_s<18.4$ & SF & $1591$ & 29 & 1.11 & $1.61\pm0.08$ & $0.306\pm0.042$ & $2.36$ & $13.71\pm3.09$  & $26.7^{+23.2}_{-13.6}$ \\
%I-K_s>4.0 EROs
$I-K_s>4.0$ & $K_s<17.9$ & E & $802$ & 85 & 1.07 & $2.11\pm0.15$ & $0.318\pm0.051$ & $1.51$ & $8.98\pm2.39$  & $7.2^{+7.6}_{-4.5}$ \\
$I-K_s>4.0$ & $K_s<17.9$ & SF & $138$ & 15 & 1.15 & $-$ & $-$ & $-$ & $-$  & $-$ \\
$I-K_s>4.0$ & $K_s<18.4$ & E & $1810$ & 68 & 1.13 & $1.96\pm0.10$ & $0.380\pm0.039$ & $3.30$ & $10.74\pm2.01$  & $12.5^{+8.3}_{-6.0}$ \\
$I-K_s>4.0$ & $K_s<18.4$ & SF & $843$ & 32 & 1.18 & $1.69\pm0.10$ & $0.377\pm0.061$ & $2.69$ & $13.77\pm3.29$  & $27.0^{+25.3}_{-14.6}$ \\
\hline
%I-[3.6]>5.0 EROs fixed gamma
$I-[3.6]>5.0$ & $[3.6]<16.5$ & E & $1849$ & 73 & 1.11 & $1.80$ & $0.568\pm0.045$ & $3.63$ & $13.36\pm1.06$  & $24.6^{+6.5}_{-5.4}$ \\
$I-[3.6]>5.0$ & $[3.6]<16.5$ & SF & $690$ & 27 & 1.09 & $1.80$ & $0.249\pm0.056$ & $8.54$ & $9.27\pm2.08$  & $7.9^{+7.0}_{-4.4}$ \\
$I-[3.6]>5.0$ & $[3.6]<17.0$ & E & $5003$ & 67 & 1.14 & $1.80$ & $0.393\pm0.025$ & $3.49$ & $11.45\pm0.73$  & $15.3^{+3.2}_{-2.7}$ \\
$I-[3.6]>5.0$ & $[3.6]<17.0$ & SF & $2461$ & 33 & 1.14 & $1.80$ & $0.176\pm0.021$ & $6.78$ & $8.22\pm0.98$  & $5.4^{+2.3}_{-1.8}$ \\
$I-[3.6]>5.0$ & $[3.6]<17.5$ & E & $9697$ & 63 & 1.16 & $1.80$ & $0.273\pm0.015$ & $1.64$ & $9.54\pm0.52$  & $8.7^{+1.6}_{-1.4}$ \\
$I-[3.6]>5.0$ & $[3.6]<17.5$ & SF & $5683$ & 37 & 1.16 & $1.80$ & $0.150\pm0.012$ & $6.93$ & $7.62\pm0.61$  & $4.3^{+1.2}_{-0.9}$ \\
%R-K_s>5.0 EROs fixed gamma
$R-K_s>5.0$ & $K_s<17.9$ & E & $1761$ & 82 & 1.04 & $1.80$ & $0.437\pm0.038$ & $3.52$ & $13.17\pm1.15$  & $23.5^{+6.7}_{-5.8}$ \\
$R-K_s>5.0$ & $K_s<17.9$ & SF & $394$ & 18 & 1.06 & $-$ & $-$ & $-$ & $-$  & $-$ \\
$R-K_s>5.0$ & $K_s<18.4$ & E & $3804$ & 71 & 1.08 & $1.80$ & $0.354\pm0.024$ & $3.13$ & $11.86\pm0.80$  & $17.0^{+3.8}_{-3.3}$ \\
$R-K_s>5.0$ & $K_s<18.4$ & SF & $1591$ & 29 & 1.11 & $1.80$ & $0.273\pm0.032$ & $3.10$ & $10.47\pm1.23$  & $11.6^{+4.7}_{-3.7}$ \\
%I-K_s>4.0 EROs
$I-K_s>4.0$ & $K_s<17.9$ & E & $802$ & 85 & 1.07 & $1.80$ & $0.342\pm0.056$ & $2.03$ & $12.69\pm2.08$  & $21.0^{+12.2}_{-8.9}$ \\
$I-K_s>4.0$ & $K_s<17.9$ & SF & $138$ & 15 & 1.15 & $-$ & $-$ & $-$ & $-$  & $-$ \\
$I-K_s>4.0$ & $K_s<18.4$ & E & $1810$ & 68 & 1.13 & $1.80$ & $0.423\pm0.037$ & $3.66$ & $13.44\pm1.18$  & $25.1^{+7.4}_{-6.2}$ \\
$I-K_s>4.0$ & $K_s<18.4$ & SF & $843$ & 32 & 1.18 & $1.80$ & $0.364\pm0.055$ & $2.87$ & $11.98\pm1.81$  & $17.6^{+9.6}_{-7.0}$ \\
\hline\hline
\end{tabular}
\\ 
Note: Clustering results for samples with $N<500$ are not shown due to poor quality statistics for our field area.\\ 
*The ERO types are labeled as E for passive EROs and SF for star-forming EROs.\\ 
\% total is the fraction of combined EROs in a particular ERO-type sub-population.\\ 
These samples contain only objects that have $z_{phot}$ information.
\end{table*}

\end{document}